\documentclass[structabstract]{aa}

\usepackage{natbib}
\usepackage{graphicx}
\usepackage{rotating}
\usepackage{txfonts}
\usepackage{lscape}
\usepackage{booktabs}
\usepackage{dcolumn}
\usepackage{float}
\usepackage{dblfloatfix}
\usepackage{array}
\usepackage{footnote}
\usepackage{wasysym}
\makesavenoteenv{tabular}
\makesavenoteenv{table}
\usepackage{diagbox}

\usepackage{color}
\usepackage{hyperref}
\hypersetup{
  colorlinks=true,        
  linkcolor=blue,         
  citecolor=cyan,         
}

\begin{document} 

\newcommand{\cf}{cf.,~}
\newcommand{\ie}{i.e.,~}
\newcommand{\eg}{e.g.,~}
\newcommand{\etal}{\textit{et al.},}
\newcommand{\lr}[1]{\textcolor{red}{LR: #1}}

\newcommand{\cmf}[1]{\textcolor{blue}{CMF: #1}}
\newcommand{\zy}[1]{\textcolor{green}{ZY: #1}}

\title{Using space-VLBI to probe gravity around Sgr A$^{*}$}
\author{C.~M.~Fromm\inst{1,2,3}, Y.~Mizuno\inst{4,1},  Z.~Younsi\inst{5,1}, H.~Olivares\inst{6,1},
O.~Porth\inst{7,1}, M.~De Laurentis\inst{8,9,1}, 
H. Falcke\inst{6,3}, M. Kramer\inst{3,10} \and L.~Rezzolla\inst{1,11,12}}

\institute{Institut f\"ur Theoretische Physik, Goethe Universit\"at, Max-von-Laue-Str. 1, D-60438 Frankfurt, Germany\
\and Black Hole Initiative at Harvard University, 20 Garden Street, Cambridge, MA 02138, USA\
\and Max-Planck-Institut f\"ur Radioastronomie, Auf dem H\"ugel 69, D-53121 Bonn, Germany\
\and Tsung-Dao Lee Institute and School of Physics and Astronomy, Shanghai Jiao Tong University, Shanghai, 200240, People's Republic of China\
\and Mullard Space Science Laboratory, University College London, Holmbury St.\,Mary, Dorking, Surrey RH5 6NT, UK \
\and Department of Astrophysics/IMAPP, Radboud University Nijmegen, P.O. Box 9010, 6500 GL Nijmegen, The Netherlands \
\and Anton Pannekoek Institute for Astronomy, University of Amsterdam, Science Park 904, 1098 XH Amsterdam, The Netherlands\
\and Dipartimento di Fisica ``E.~Pancini", Universit\'a di Napoli ``Federico II", Via Cinthia, I-80126, Napoli, Italy\
\and INFN Sez.~di Napoli, Via Cinthia, I-80126, Napoli, Italy \
\and Jodrell Bank Centre for Astrophysics, University of Manchester, Machester M13 9PL, UK \
\and Frankfurt Institute for Advanced Studies, Ruth-Moufang-Strasse 1, 60438 Frankfurt, Germany \
\and School of Mathematics, Trinity College, Dublin 2, Ireland \\
\email{cfromm@th.physik.uni-frankfurt.de}}

   \date{
Draft 1.0: \today
}

  \abstract
{The Event Horizon Telescope (EHT) will soon provide the first high-resolution images of 
the Galactic Centre supermassive black hole (SMBH) candidate Sagittarius A* (Sgr A$^{*}$),
enabling us to probe gravity in the strong-field regime.
Besides studying the accretion process in extreme environments, the obtained data and reconstructed 
images could be used to investigate the underlying spacetime structure. In its current configuration, the EHT
is able to distinguish between a rotating Kerr black hole and a horizon-less object like a 
boson star. Future developments can increase the ability of the EHT to tell different spacetimes apart.
}
{ 
We investigate the capability of an advanced EHT concept, including an 
orbiting space antenna, to image and distinguish different spacetimes around Sgr A$^{*}$.
}
{We use general-relativistic magneto-hydrodynamical (GRMHD) simulations of 
accreting compact objects (Kerr and dilaton black holes, as well as boson stars) and 
compute their radiative signatures via general-relativistic radiative transfer (GRRT). 
{To facilitate comparison with upcoming and future EHT observations we produce realistic synthetic
data including the source variability, diffractive and refractive scattering while incorporating the observing array, including a space antenna. From the generated synthetic observations we dynamically reconstructed black hole shadow images using regularised Maximum Entropy methods. We employ a genetic algorithm to optimise the orbit of the space antenna with respect to improved imaging capabilities and {u-v-plane coverage} of the combined array
(ground array and space antenna and developed a new method to probe the source variability in Fourier space.}}
{{The inclusion of an orbiting space antenna improves the capability of the EHT to 
distinguish the spin of Kerr black holes and dilaton black holes based on reconstructed radio images and complex visibilities.}}
   {}

   \keywords{Physical data and processes: Gravitation, Magnetohydrodynamics (MHD),
radiation mechanisms: thermal --
Methods: numerical --
Galaxies: individual: Sgr A$^{*}$ --
Techniques: interferometric
               }

\titlerunning{Using space-VLBI to test spacetimes around Sgr A$^{*}$.}
\authorrunning{C.~M.~Fromm et al.}
\maketitle
\section{Introduction}
\label{intro}
The Event Horizon Telescope presented the first {horizon scale} images of the black hole
in M87 \citep{2019ApJ...875L...1E} and will soon provide the first image of the black hole candidate Sgr A$^{*}$ in our Galaxy. The current configuration of the EHT consists of eight telescope scattered across Europe, 
North- and South America and the South Pole \citep{2019ApJ...875L...2E}. Combining 
the data recorded simultaneously by the individual telescopes after the observations, 
$\mu$as angular resolution is achieved. These observations enable us, for the first time, 
to study accretion processes in M87 and in the centre of our galaxy with unparalleled resolution and 
to probe gravity in the strong-field regime. Given the small number of telescopes 
participating in the observations, the intrinsic variability of the source and the {interstellar
scattering screen}, reconstructing an image and {discriminating among different spacetimes}
is challenging \citep{2014ApJ...788..120L,2016ApJ...817..173L,2018NatAs...2..585M}. 
An interferometer such as the EHT samples the brightness distribution of an astronomical 
object in Fourier-space: the so called \emph{u-v plane}. Due the limited number of participating
telescopes, the u-v plane is sparsely sampled. The sampling of this u-v plane increases
with the number of telescopes and the duration of the observations. However, an
improvement in the resolution of the image can be only be obtained by increasing the
distance between the telescopes, the so-called baselines, or by increasing the observed
frequency. Given the current configuration of the EHT array, a significant increase in the
baselines can only be achieved by extending baselines to space. Using the ground
array while increasing the observed frequency {above 345\,GHz} is limited by the opacity of the atmosphere
and its water vapour content.

The concept of space-based Very Long Baseline Interferometry (VLBI) including a ground array has
been studied extensively since the early 1970s \citep[see][for a historic overview]{2012rsri.confE..10S}
and successfully launched space antennas are Highly Advanced Laboratory for Communications and Astronomy (HALCA)  (project VLBI Space Observatory Programme, (VSOP)) \citep{1998Sci...281.1825H} and,
more recently, Spektr-R (project RadioAstron) \citep{2013ARep...57..153K}. Both missions operate at lower
frequencies than the EHT (230\,GHz) and in the case of RadioaAstron provide $\rm \mu as$
resolution \citep{2016ApJ...817...96G}. Therefore, in this work we consider the increase {in} the
angular resolution that can be obtained by extending the baselines of the EHT via a space-based antenna \citep[see also][]{2019ApJ...881...62P}.
This configuration has the advantage of using a well-calibrated ground array. For a mission based 
entirely on space telescopes, see a recent studies by \citet{2019A&A...625A.124R,2019arXiv190309539F}.
Within this work we address the scientific question as to whether such a configuration will
improve the current ability of the EHT to distinguish {among} different theories of 
gravity using radio observations of Sgr A$^{*}$.
\newline The structure of this paper is as follows: in Sect.~\ref{GRMHD} we briefly introduce our 
GRMHD simulations and GRRT calculations.
The procedure for the selection of the orbit of the space antenna is introduced in Sect.~\ref{GA} 
and the results of the synthetic imaging and data analysis are shown in Sect.~\ref{res}. 
We present our discussion and conclusions in Sect.~\ref{dis}.
Sgr A$^{*}$ is located at RA: $17^h\,45^m\,40.0409^s$
and DEC:$-29^\circ\,45^\prime\, 40.0409^{\prime\prime}$, at a distance of {$\rm 8178\pm 13\,stat.\pm 22\,sys.\,pc$ } and
its central candidate SMBH exhibits a mass of $M_{\rm bh}=4.14\times10^6M_{\astrosun}$ \citep{2019A&A...625L..10G}.
\section{GRMHD and GRRT simulations}
\label{GRMHD}
We use the state-of-the-art GRMHD code \texttt{BHAC} \citep{2017ComAC...4....1P,2019A&A...629A..61O} and perform three-dimensional 
accretion simulations on to Kerr black holes (${\rm a}_{*}$=0.6 and ${\rm a}_{*}$={0.94}), dilaton black holes and a 
boson star \citep{2018NatAs...2..585M,2020MNRAS.497..521O}.
Here ${\rm a}_{*}$ is the dimensionless black-hole spin parameter, with $|\rm a_{*}|<1$.

\begin{figure*}[t]
\resizebox{\hsize}{!}{\includegraphics{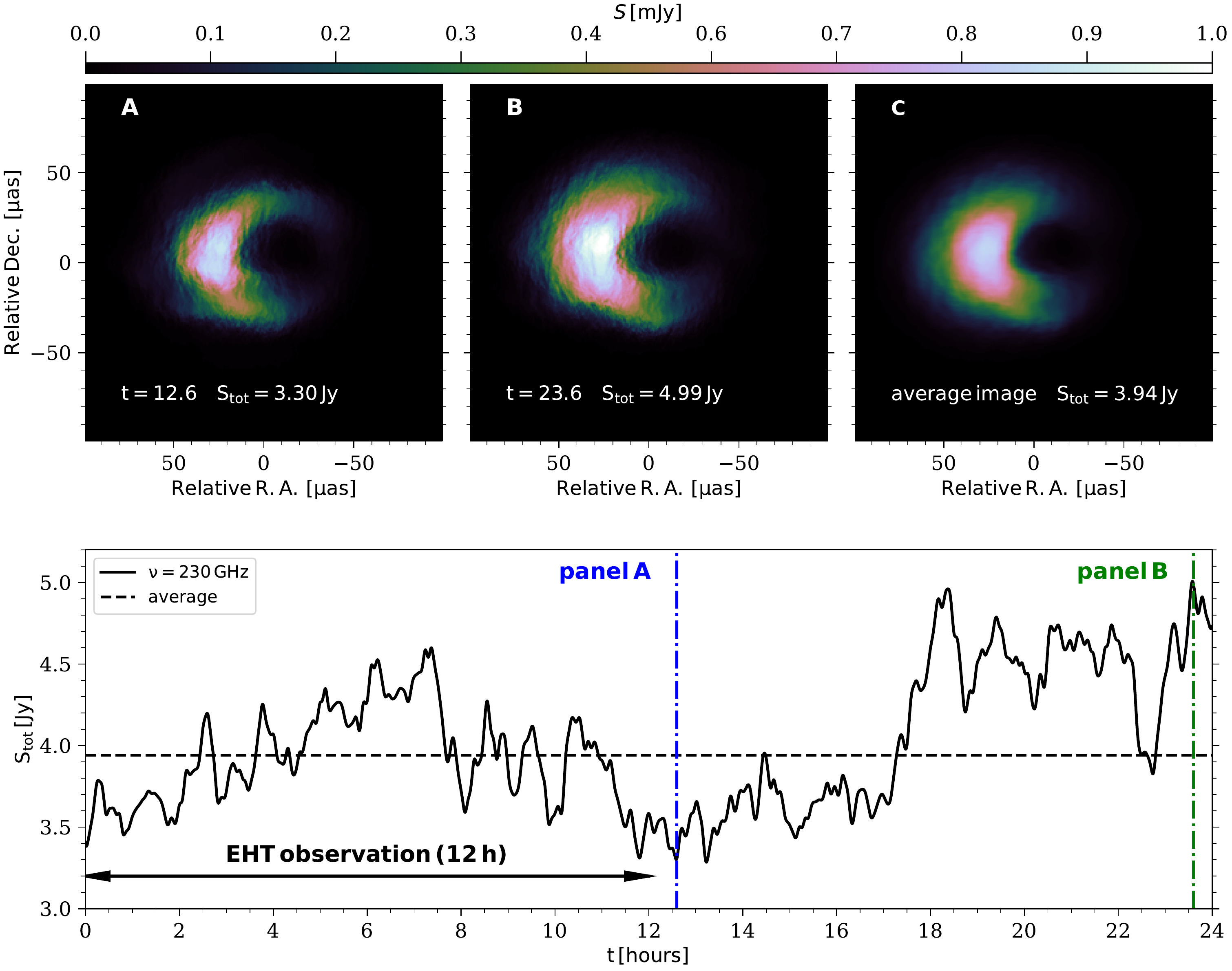}} 
\caption{Top: panels A and B show GRRT images for a Kerr black hole with ${\rm a}_{*}$=0.6  including scattering effects ( diffractive and refractive scattering) for two selected times t=12.6\,h and t=23.6\,h (dash-dotted line in the bottom panel) at a frequency of 230\,GHz. The average image for 24\,hours of observation is presented in panel C.  Bottom: simulated total flux light curve at 230\,GHz for 24\,hours of observations. Indicated is a typical EHT observation with 12 hours duration.}
\label{fluxvar} 
\vspace{-12pt} 
\end{figure*}

We have chosen a black hole solution in the Einstein-Maxwell-dilaton-axion (EMDA) theory, which derives from a particular string theory \citep{1994PThPh..92...47O,1995PhRvL..74.1276G}. Such black-hole solutions exhibit more diverse physical properties than others in the literature. For example, they have hair, in addition to mass, rotation and charge, so as to be an interesting laboratory for performing black hole experiments and studying possible differences between Einstein's gravity and alternative theories of gravity.
Moreover, since both the dilaton and the axion are considered to be candidates for dark matter, the study of the shadow from a dilaton black hole may provide hints on the observational properties of the matter in which the black hole is immersed \citep{PhysRevLett.123.021102}.

We also consider a stable boson star, which although classified as a compact object, is physically completely different from a black hole \citep{1997PhLB..404...25S}. It is well known that the gravitational field of a boson star bends light around itself, creating a region \emph{resembling} the shadow of a black hole's event horizon. Like a black hole, a boson star will accrete ordinary matter from its surroundings, but its opacity means this matter (which likely would heat up and emit radiation) would be visible at its center. There is no significant evidence so far that such stars exist. However, it may become possible to detect them through VLBI measurements \citep{2020MNRAS.497..521O}. In theory, a super-massive boson star could exist at the core of a galaxy, which might explain many of the observed properties of active galactic nuclei. Boson stars have also been proposed as candidate dark matter objects, and it has even been hypothesised that the dark matter haloes surrounding most galaxies might be viewed as enormous boson stars \citep{PhysRevLett.121.151301}.

The formation of boson stars or other exotic objects is an interesting process to study. Although we know that at the center of our galaxy there is a highly compact object, it is important to study different candidates because we might find that other galaxies could harbour exotic objects composed of ``non-baryonic'' matter at their centres \citep{1996PhRvD..53.2236L}.

For spacetimes not described by Einstein's general theory of relativity, we use the 
Rezzolla-Zhidenko metric parameterisation \citep{2014PhRvD..90h4009R,2016PhRvD..94h4025Y}. The initial conditions 
of all GRMHD simulations for Kerr and dilaton black holes as well as for the boson star, consist of a torus in hydrodynamical equilibrium with a weak 
poloidal magnetic field. In order to trigger the accretion process the magnetic field in the 
torus is seeded with a small perturbation which leads to the formation of the magneto-rotational 
instability (MRI). After the saturation of the MRI (${\rm t} > 5000\,\mathrm{GM/c^2}$, with gravitational constant, G, mass of the central object, M, and speed of light c) all simulations 
show a quasi-stationary accretion flow \citep[for more details see][]{2018NatAs...2..585M,2020MNRAS.497..521O}.

In the next step we compute the radiative signature of the accretion process via GRRT 
calculations using the \texttt{BHOSS} code {\citep{2020IAUS..342....9Y,2020ApJ...897..148G}}. We use an viewing angle of $\theta=60^\circ$, adjust the ion-to-electron 
temperature ratio ($T_i/T_e=3$) and the mass-accretion rate to in order to 
{obtain the  an average flux of 4 Jy at 230 GHz \citep[see {supplemental information} in][for a detailed 
description of the GRRT calculations and fitting procedure]{2018NatAs...2..585M}. This leads to a flux density variation between 3 Jy and 5 Jy  (see light curve in the bottom panel of Fig. ref{fluxvar}) which is in agreement with the observed flux density variations 2 Jy -- 5.5 Jy provided by \citet{2015ApJ...802...69B}}
 The GRRT images are computed every 10\,M which corresponds to 200\,s for SgrA* for a time span of 4320\,M (24\,h for SgrA). This long duration allows for the calculation of several overlapping and two independent 12\,h\footnote{typical duration for EHT observations of SgrA$^*$ which allows European and American telescopes to participate in the observation.} observational window.
In this work we investigate  the following spacetimes: Kerr black holes in general relativity,
a dilaton black hole and a boson star.
The variability of the total flux and in the emission structure of the individual GRRT snapshots,
together with the averaged image, is presented in Fig.~\ref{fluxvar} and in Table~\ref{GRRTinfo} we present an overview of the GRRT images used.
\begin{table}[h!]
\vspace{-0pt}
\centering
\caption{Overview of simulations used}
\begin{tabular}{lllll} 
\hline\hline
Spacetime  & ${\rm a}_{*}$ & $\theta$  [deg] & $\Delta t$ [M (s)] & t [M (h)]\\
\hline
Kerr	&0.60	&60 	& 10 (200) & 4320 (24)\\
Kerr	&0.94	&60	&10 (200) & 4320 (24)\\
dilaton	&0	&60 	& 10 (200)& 4320 (24)\\
boson star	&0	&60& 10 (200)& 4320 (24)	\\
\hline
\label{GRRTinfo}
\end{tabular} 
\vspace{-12pt} 
\end{table} 

\section{Orbit selection and optimisation strategy}
\label{GA}
{In this section of  our exploratory paper we present one possible array optimisation strategy which could lead to improved imaging capabilities for the EHT and thus may allow us to distinguish among different spacetimes.}
Thus, the question arises as to which kind of orbit is required for improving the imaging of the Galactic Centre.
The answer to this question depends strongly on the observational constraints: the position of 
the Galactic Centre relative to the space and ground antennas and the duration of the 
observations. {In the following we will assume an observation schedule consisting of 12 hours of 
Sgr A$^{*}$ observations. This long observing time allows European as well as North- and South American telescopes to participate in the observation. }

For the space-EHT concept we consider the EHT 2017 
configuration as the ground array and include the Northern Extended Millimeter Array (NOEMA\footnote{\url{http://iram-institute.org/EN/}}) and a space antenna.
In Table \ref{SEFD} we list each antenna and its corresponding system 
equivalent flux density (SEFD) \citep{2019ApJ...875L...2E}.
\begin{table}
\vspace{-0pt}
\centering
\caption{Effective antenna diameter $d$ (for single telescopes this corresponds to the diameter of the dish) and system equivalent flux density (SEFD) used for the ground and space antennas {\citep[see][for details]{2019ApJ...875L...2E}}}
\begin{tabular}{lll} 
\hline\hline
Telescope  & $d_{\mathrm{eff}}$ [m] & SEFD [Jy]\\
\hline
ALMA$^a$	&73	&74 	\\
APEX	&12	&4700	\\
JCMT	&15	&10500 	\\
LMT	&32.5	&4500 	\\
NOEMA$^\ast$	&52 	&700	\\
PV	&30	&1900	\\
SMT	&10   	&17100	\\
SMA$^b$	&14.7  	&6200	\\
SPT$^c$	&6  	&19300	\\
Space Antenna$^\ast$ & 8 & 20000\\
\hline
\multicolumn{3}{l}{$^\ast$ only used for the space-EHT configuration.}\\
\multicolumn{3}{l}{$^a$ for EHT\,2017 ALMA used 37 $\times$ 12 m.}\\
\multicolumn{3}{l}{$^b$ for EHT\,2017 SMA used 6 $\times$ 6 m.}\\
\multicolumn{3}{l}{{$^c$ for EHT\,2017 SPT was under-illuminated}}\\
\multicolumn{3}{l}{{with an effective diameter of 6 m.}}
\label{SEFD}
\end{tabular} 
\vspace{-12pt} 
\end{table} 
For the orbiting space antenna we assume a diameter of 8 meters (similar to the size of Spektr-R), a system temperature 
of 100\,K and an antenna efficiency of 40\%. Given these values, a SEFD of $20\times10^3$\,Jy 
is computed. The orbit of the space antenna can be described by six orbital elements which 
can be divided into orbital shape parameters (semi-major axis, $a$, and eccentricity, $e$) and
orbital orientation parameters which provide the location and orientation of the orbit in space relative to the 
equatorial plane (inclination, $i$, longitude of the ascending node, $\Omega$, argument of 
perigee, $\omega$, and the true anomaly, $\vartheta$). The six orbital elements of the space antenna
together with duration of the observations lead to a seven-dimensional parameter space which 
has to be searched to obtain an optimal configuration. This task can be formulated as a 
constrained non-linear optimisation problem and written as follows:
\begin{eqnarray}
\begin{array}{ll@{}ll}
\mathrm{minimize:}  & \displaystyle  f(\vec{x}) \,, & \\
\mathrm{subject \ to:}& \displaystyle g_{j}(\vec{x})\leq 0\,, &   & j=1 ,..., n\,, \\
                 &  \displaystyle        x_{L,i}\leq x_{i}\leq x_{R,i}\,,  & & i=1 ,..., m\,,
\end{array}
\end{eqnarray}
where $\vec{x}$ is a seven-dimensional vector containing the model parameters, i.e.,
$\vec{x}\equiv\left[t_\mathrm{obs}, a, e, i, \Omega, \omega,\vartheta \right]^{T}$, $f(\vec{x})$ is the 
objective function (minimisation function), $g_j(\vec{x})$, are the constraints and $x_{L,i}$ 
and $x_{R,i}$ are lower and upper boundaries for the model parameters. In Table \ref{parabound} we report the boundaries used during the optimisation.\\

\begin{table}[h!]
\vspace{-0pt}
\centering
\caption{{Parameter boundaries used for the optimisation}}
\begin{tabular}{@{}lllllll@{}} 
\hline\hline
\small
 $t_{\rm obs}$ [h] & $a$ [$10^3$km] & $e$ & $i$ [deg] & $\Omega$ [deg] & $\omega$ [deg] & $\vartheta$ [deg]\\
\hline
0 -12 & 0.4-100 & 0-0.9 & 0-180& 0-360 & 0-360 & 0-360 \\
\hline
\label{parabound}
\end{tabular} 
\vspace{-12pt} 
\end{table}

{Improving the imaging capabilities of the EHT can be translated into an increased angular resolution
and a denser sampling of the u-v plane as compared to the current 
EHT configuration. The {improvement of} the imaging capabilities {\ie better array resolution and u-v plane coverage} can be {assessed} by computing image metrics between the infinite resolution GRRT image and the reconstructed {image}. 
Therefore, we use for the minimisation process a combination of the denser sampling in the u-v plane (minimising the distance between u-v points) and the improved imaging (improved image metrics). For the image metrics we used {the} normalised cross correlation coefficient (nCCC) and the structural dissimilarity measure (DSSIM)\footnote{smaller values indicate a better image agreement} \citep{2004ITIP...13..600W}. The minimisation function is given by:}
\begin{equation}
f\left(\vec{x}\right)=\mathrm{DSSIM}\left(\vec{x}\right),
\label{f1} 
\end{equation}
and we use the following constraints\footnote{the constraints are fulfilled if the $g_i<0$} to speed up the optimisation procedure:
\begin{eqnarray}
g_1&=&\mathrm{DSSIM}\left(\vec{x}\right)- 0.75\times\mathrm{DSSIM_{EHT2017}}, \\
\label{ccconst1}
g_2&=&\Delta r_{\rm uv,\,max}\left(\vec{x}\right)-\Delta r_{\rm uv,\,max}^{\rm EHT2017} \,.
\label{ccconst2}
\end{eqnarray}
{The first constraint ensures that the DSSIM is improved by at least 25\% as compared to image reconstructed using the EHT 2017 configuration. The denser sampling of the u-v plane is addressed by the second constraint, $g_2$, where we enforce that the projected distance between the u-v points, $\Delta r$, is minimised with respect to u-v sampling of the EHT 2017 configuration.\\
During the optimisation we generate for each set of parameters \eg for each $\left(\vec{x}\right)$ synthetic visibilities including both, diffractive and refractive scattering \citep{2016ApJ...833...74J} and reconstruct the image using the \texttt{EHTim} package
\footnote{https://github.com/achael/eht-imaging} \citep{2016ApJ...829...11C,2018ApJ...857...23C}.}

{Due to the numerical costs we use during the optimisation a single GRRT image instead of a series of GRRT images (or GRRT movie) and the array configuration and additional parameters used during the data generation and image reconstruction are listed in Table \ref{SEFD} and \ref{reconst}. The procedure of the image reconstruction follows the approach of \citet{2016ApJ...829...11C}: initialisation of the imaging using as gaussian prior with a FWHM of 70\,$\mu$as and the repeated re-initialisation of the imaging using the previously obtained image convolved with half the nominal array resolution. During the orbit optimisation we use two re-initialisation loops and we align the reconstructed image with the GRRT image prior to the computation of the DSSIM.}
\begin{table}[h!]
\vspace{-0pt}
\centering
\caption{Parameters used for the data generation and image reconstruction}
\begin{tabular}{lllll} 
\hline \hline
\multicolumn{5}{c}{synthetic data generation}\\
$t_{\rm int}$ [s] & $\Delta \nu$ [GHz] & $t_{\rm gap}^a$ [s] & $\nu_{\rm obs}$ [GHz] & gain off set$^b$\\
\hline
12	&4, 8$^\ast$	&200 	& 230 & 0.1\\
\hline 
\multicolumn{5}{c}{image reconstruction}\\
data &  weighting$^c$ & \multicolumn{2}{c}{regularizer}  &  weighting$^c$ \\
\hline
visibilities	&100	&\multicolumn{2}{c}{simple entropy}	&2 \\
bi-spectra	&10	&\multicolumn{2}{c}{simple entropy} 	& 1\\
\hline
\multicolumn{5}{l}{$^\ast$ only used for the space-EHT configuration.}\\
\multicolumn{5}{l}{{$^a$ time difference between GRRT images}}\\
\multicolumn{5}{l}{{$^b$ antenna gains are drawn form a normal distribution with}}\\
\multicolumn{5}{l}{{mean=1.0 and standard deviation=0.1 (gain off set).}}\\
\multicolumn{5}{l}{{$^c$ weighting factor for data terms and image entropies}}\\
\multicolumn{5}{l}{{used during the image reconstruction.}}
\label{reconst}
\end{tabular} 
\vspace{-12pt} 
\end{table} 
\begin{figure*}[t!]
\resizebox{\hsize}{!}{\includegraphics{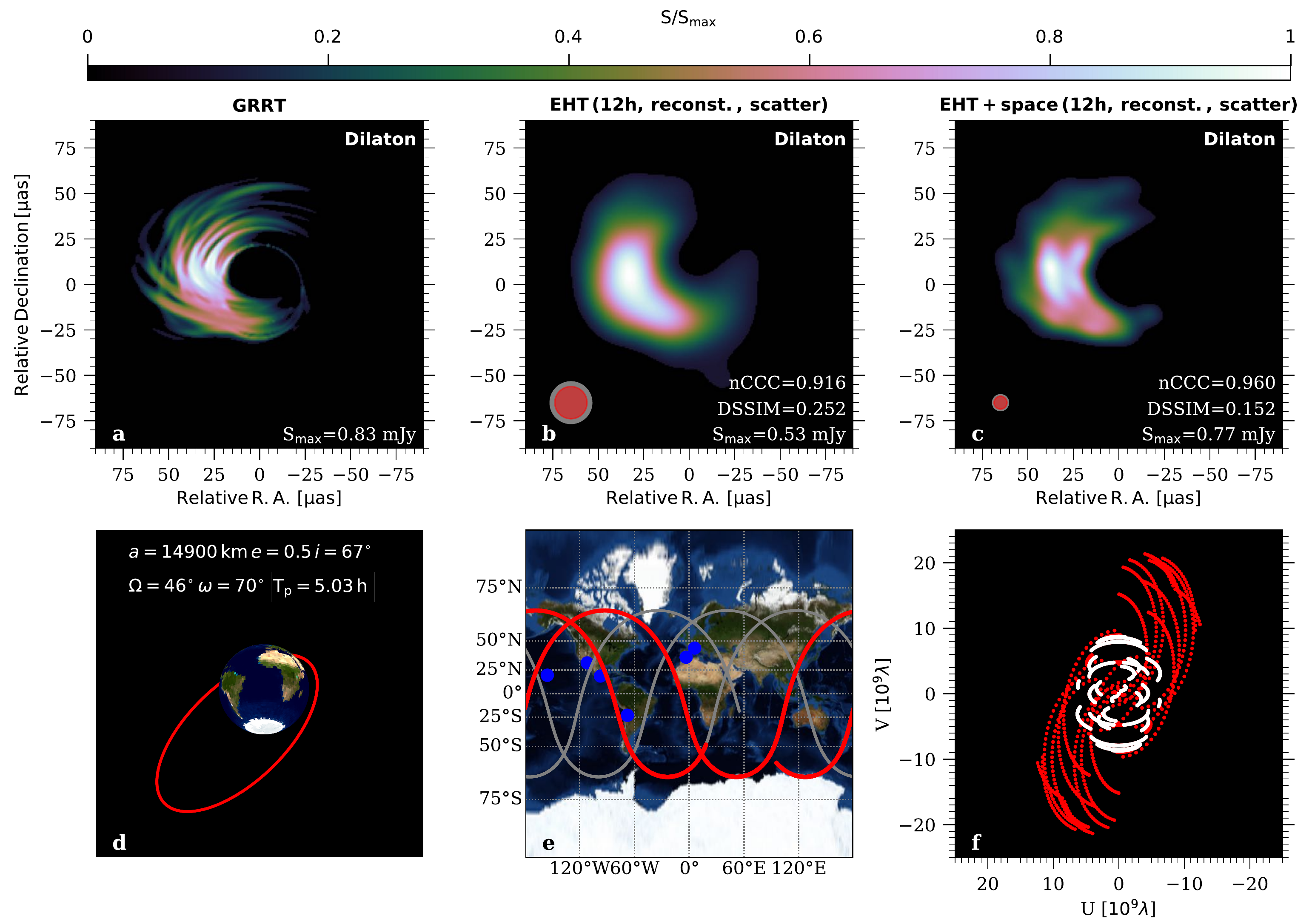}} 
\caption{Result of the orbit optimisation for 12\,hours of Sgr A$^{*}$ observations. 
Top row from left to right: GRRT image (panel a), reconstructed 
image with interstellar scattering (including both, diffractive and refractive scattering during the generation of the synthetic visibilities) using the EHT 2017 configuration convolved with 75\% (red 
shading) of the nominal array resolution (light grey shading, panel b) and reconstructed 
image with interstellar scattering (including both, diffractive and refractive scattering during the generation of the synthetic visibilities) using the space-EHT configuration convolved with 75\% (red shading) of the nominal array resolution (light grey shading, panel c) for a dilaton black hole. Bottom row from left to right: satellite orbit as seen from SgrA$^\star$ {with orbital parameters and orbital period} (panel d), satellite ground track (red lines for 12\,hours and grey ones for 24\,hours) and ground array antennas (blue points, panel e) and u-v sampling for 
the ground array (white points) and the baselines including the 
space antenna (red points, panel f).} 
\label{orbitres} 
\vspace{-12pt} 
\end{figure*}
Given the high dimensionality of this constrained optimisation problem, gradient-based solvers may become 
stuck in a local minimum and/or required large computational resources to map out the 
gradient of the parameter space with sufficient resolution to avoid this problem. An elegant method to 
circumvent the above mentioned difficulties is based on gradient-free optimisation algorithms. 
In this work we apply a genetic algorithm (GA) and in particular employ the 
implementation of a Non Sorting Genetic Algorithm II (\texttt{NSGA2}) to solve the 
optimisation problem \citep{Deb:2002}. For the initial generation we use 1000 randomly 
initiated orbits and we evolve them for 100 generations.\\

\begin{figure}[h!]
\resizebox{8.8cm}{!}{\includegraphics{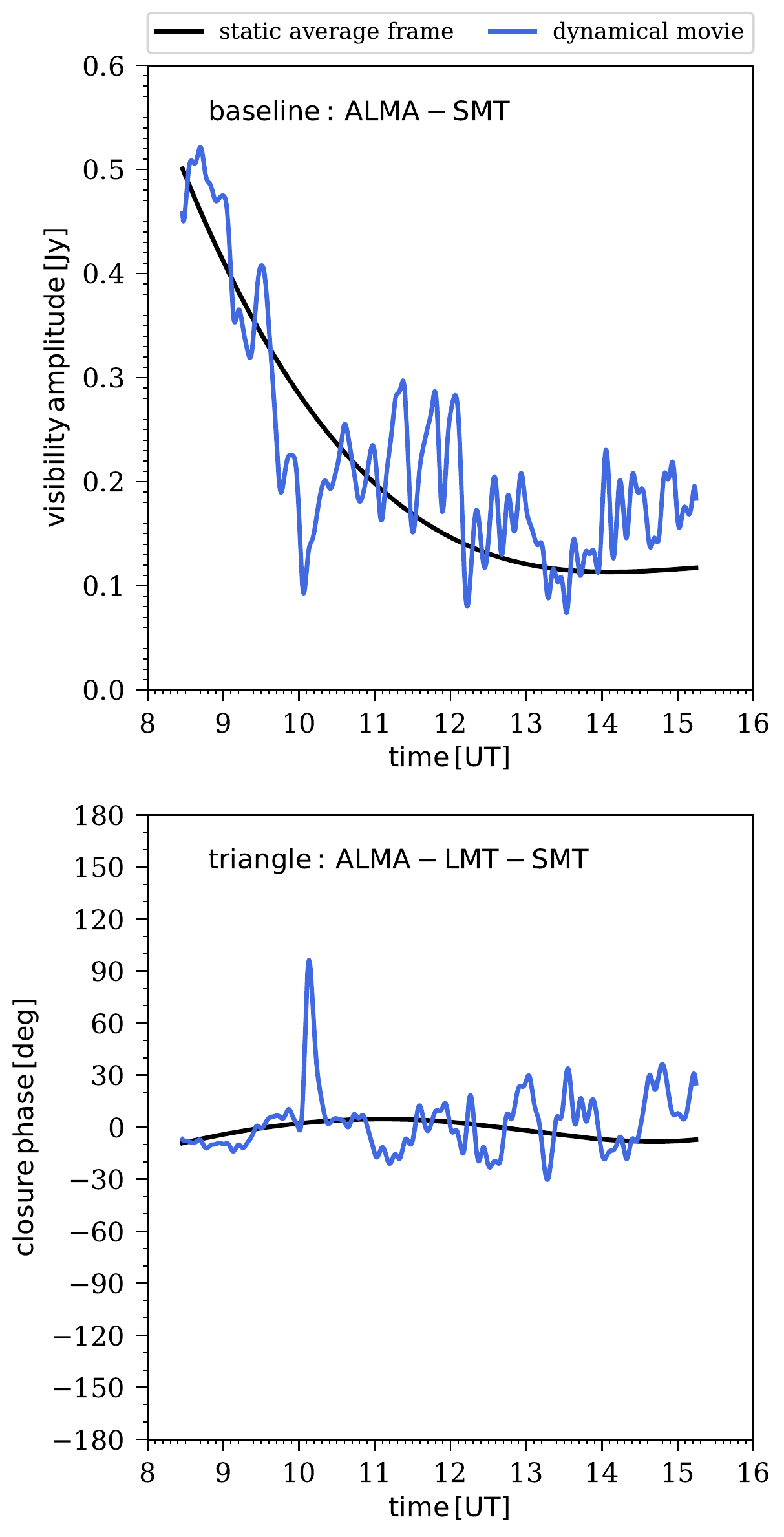}} 
\caption{Comparison between synthetic data generation from a dynamical movie (blue lines) and from its static average frame (black lines) for the visibility amplitude (top) and closure phase (bottom) for Kerr black hole with ${\rm a}_{*}$=0.6.} 
\label{vacpvar} 
\end{figure}

{The results of our numerical optimisation are presented in Table~\ref{resobs} and Fig.~\ref{orbitres}.
The improvement of the image reconstruction is clearly visible by comparing the EHT\,2017 reconstructed image (panel b in Fig.~\ref{orbitres}) with the image obtained by space-EHT (panel c  in Fig.~\ref{orbitres}). The ground truth GRRT image is presented in panel a in Fig.~\ref{orbitres}. The space-EHT image is able to capture and image fine flux arcs which are smeared out in the EHT 2017 configuration. The visible image improvements are also reflected {in} the improved image metrics: the nCCC increased from 0.92 to 0.96 and the DSSIM decreased by 40\% from 0.252 to 0.152 (see also constraint $g_1$ Eq. \ref{ccconst1}). Notice that increased nCCC and decreased DSSIM corresponds to a better image matching. The satellite orbit as seen from Sgr A$^\star$  (note that the Earth is viewed from $-30^\circ$) is presented in panel d in Fig.~\ref{orbitres}. In panel e in Fig.~\ref{orbitres} the satellite ground track (projection of the satellite orbit onto
the surface of the Earth) of the space antenna (red points for 12\,hours and grey ones for 24\,hours) 
and the ground antennas are indicated by the blue points. The sampling of the u-v plane is presented in panel f in Fig.~\ref{orbitres} where the white points indicate the u-v tracks of the ground array and the red points the u-v tracks including the space antenna. The addition of the space antenna is not only extending the u-v sampling up to 25\,G$\lambda$ but also adds short and intermediate baselines to the array as required by the constraint $g_2$  (see Eq. \ref{ccconst2}) of the optimisation process which also improve the imaging capabilities.}\\
\setlength{\tabcolsep}{0.5em}
\begin{table}[h!]
\centering
\caption{Optimised satellite orbit for 12 hours of Sgr A$^{*}$ observation.}
\label{resobs}
\begin{tabular}{@{}c c c c c c c c@{}}
\hline\hline
a [km]& i [$^\circ$] & e & $\Omega$  [$^\circ$] & $\omega$  [$^\circ$]& $\vartheta$  [$^\circ$] & $t_\mathrm{obs,start/stop}$ [UT] & {$\mathbf{T_p}$ [h]}\\
\hline
14900 & 67 & 0.5 & 46 & 70 &330 & 04:00 -- 16:00  &{ 5.03}\\
\hline 
\end{tabular}
\vspace{-12pt} 
\end{table} 
\newline {The orbit optimisation procedure suggested an elliptical orbit with a semi-major axis of $a=14900$\,km with an eccentricity of $e=0.5$ at an inclination of 67$^\circ$. The optimal time span of the SgrA$^\star$ observation is obtained between 4:00 UT and 16:00 UT. {The orbital period of the satellite is $\rm T_p=5.03$\,h. Notice, that elliptical orbits with long orbital periods have also been used for VSOP ($e$=0.6, $\rm T_p=6.3$\,h) and the Spekt-R ($e=0.9$, $\rm T_p$=200\,h) \citep{1998Sci...281.1825H,2013ARep...57..153K}. In Appendix \ref{orbittest} we illustrate the impact of different satellite orbits on the image reconstruction.} Our results differ from the results of \citet{2019ApJ...881...62P} which used a circular orbit with semi-major axis $a=6652$\,km at an inclination of 61$^\circ$. The similarity in the inclination is due to the declination (DEC) of SgrA$^\star$ of -29$^\circ$ and a face-on orbit is given by a inclination $\sim \mathrm{DEC}\pm90^\circ$. The difference in the eccentricity, $e$, and  semi-major axis, $a$, can be explained by the different models used for SgrA$^\star$. \citet{2019ApJ...881...62P} uses a Schwarzschild black hole (a=0) at an inclination of 10$^\circ$. In this work we use four different physical models: GR black holes (a=0.6 and a=0.94), a Dilaton black hole and a boson star all at an inclination of $60^\circ$. Due to the larger relativistic effects at higher inclination (Doppler boosting) the GRRT images show a large left-right asymmetry \eg the flux distribution is more compact as compared to black holes seen at an inclination of $10^\circ$. In addition the boson star used in our work shows the most compact flux distribution around $10\,\mathrm{\mu as}$ (see panel i in Fig. \ref{finalimage}). Since we require in our optimisation  to resolve and recover the compact structure of the black holes and at the same time the boson star we need to improve the angular resolution (longer base lines) and the u-v plane coverage (short baselines). Using one satellite this can be achieved by an elliptical orbit. In our case the perigee is located at a height of $\sim1100\,\mathrm{km}$ and the apogee at a height of $\sim 22300\,\mathrm{km}$. The space-ground tracks of this orbit are presented in panel f of Fig. \ref{orbitres} and the plot shows clearly the improved u-v coverage (short baselines with $\sqrt{u^2+v^2}<9\,G\lambda$ and long base lines $\sqrt{u^2+v^2}>9\,G\lambda$). \\
\noindent Using this satellite orbit and the observing time span from 4:00-16:00 UT the reconstructed images for all black holes and the boson star clearly improve the quality of the reconstructed images (better image metrics as compared to the EHT 2017 array) and are able to recover the structure seen in their infinite resolution GRRT counterparts (see Fig. \ref{finalimage}).}\\
\newline {After the optimisation of the space-EHT \eg defining the orbit of the space antenna and the observing time we create synthetic data taking the source variability during the course of the EHT observation into account. Therefore we create for each of the space-times under investigation a movie from the GRRT images which covers 12\,h of observations (first 216 GRRT images are used, see indicated time 12\,h time span in Fig. \ref{fluxvar}). From this movie we create the synthetic visibilities using the same parameters as used during the orbit optimisation (see Table \ref{SEFD} and \ref{reconst}).  As a consequence of creating the synthetic data from a variable source the visibility amplitude (VA) of the zero baselines (ALMA-APEX, JCMT-SMA) is variable and not stationary as in the case of using an 12\,h  averaged GRRT image for the data creation. This behaviour is also true for non-zero baselines and for the closure phases (CP). In Fig. \ref{vacpvar} we compare the VA and CP created from GRRT movie and from its static average frame for the ALMA-SMT baseline and for {the} ALMA-LMT-SMT triangle. The differences in the behaviour of the VA and CP are clearly visible and {the} most striking difference can be seen at t=10\,UT.}
\section{Results}
\label{res}
{Given the underlying variability in the flux density and also in source size (see Fig. \ref{fluxvar}) of the different models, standard VLBI imaging approaches which assume a static source during the course of the observations are likely to fail reconstructing an image. Therefore we follow the approach of \citet{2017ApJ...850..172J} and use dynamical imaging to reconstruct an average image from the time variable data. Since we are mainly interested in the average image obtained during the observation and our GRRT images do not show significant flaring events\footnote{we define a flare as doubling of the flux density from one frame to another} we use the $\mathcal{R}_{\Delta I}$ regularizer with $D_2$ distance metric and complex visibilities as data product \citep[see][for more details]{2017ApJ...850..172J}. We reconstructed 24 frames with a duration of 30 minutes each for a total duration of 12\,h. For the first initialisation of the imaging we applied a circular gaussian with FWHM=70\,$\mu$as prior image and for repeated re-initialisation of the imaging we use as prior the previously obtained average image computed from all 24 frames convolved with the half of the nominal array resolution. For the dynamical reconstruction we applied 5 re-initialisation loops.}
\subsection{Image Plane comparison}
\begin{figure*}[h!]
\centering
\vspace{-10pt}
\resizebox{0.95\hsize}{!}{\includegraphics[]{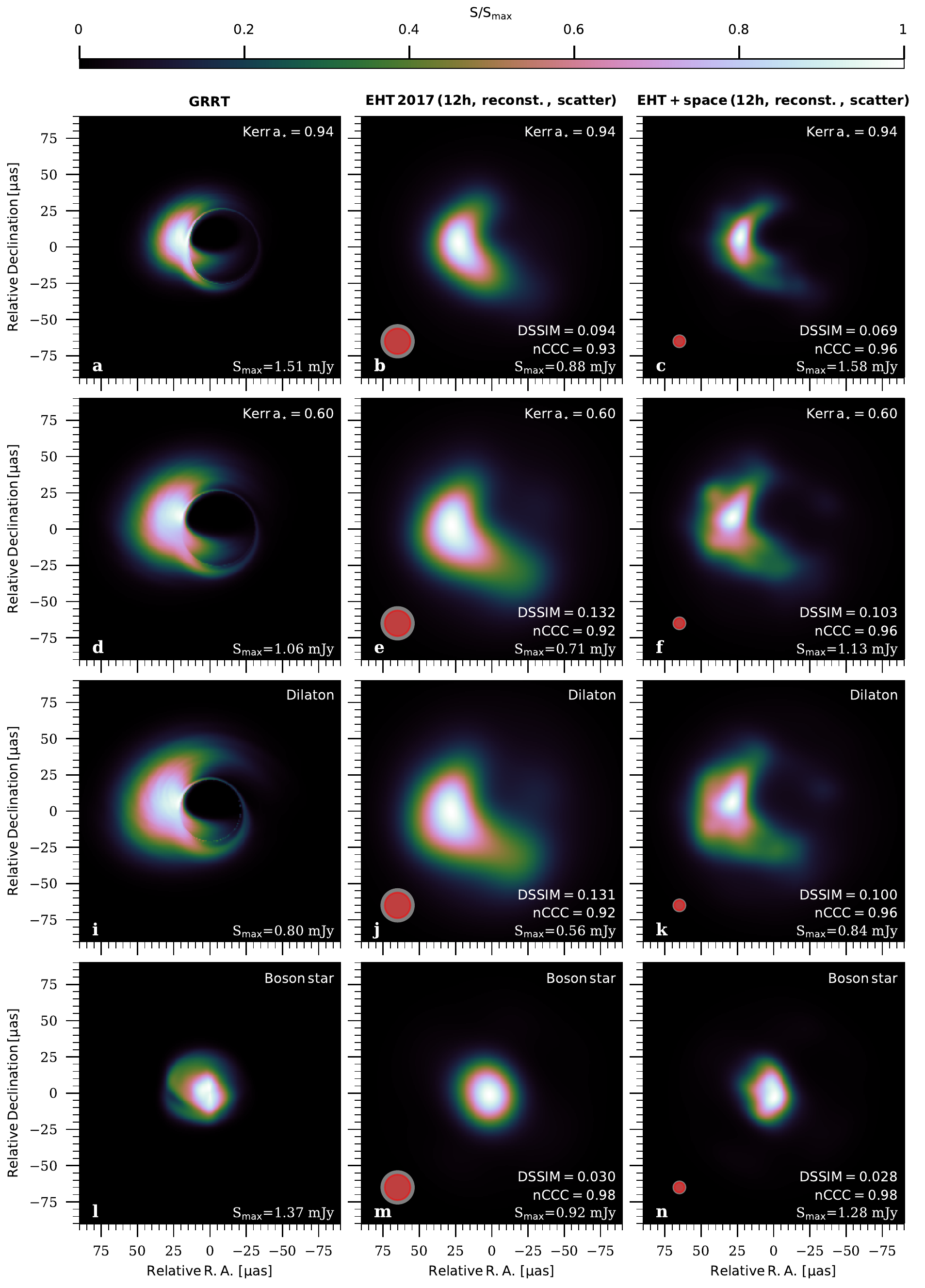}} 
\caption{Synthetic black hole images for Sgr A$^{*}$ for a Kerr black hole with ${\rm a}_{*}$=0.94 (panels a-c),
a Kerr black hole with ${\rm a}_{*}$=0.6 (panels d-f), a non-rotating dilaton black hole (panels g-i) and for a stable, non-rotating boson star 
(panels j-l). From left to right and for all rows: GRRT image (panels a, d, g, j), dynamically reconstructed 
image with interstellar scattering (including both, diffractive and refractive scattering during the generation of the synthetic visibilities) using the EHT 2017 configuration convolved with 75\% (red 
shading) of the nominal beam size (light grey shading, panels b, e, h, k) and dynamically reconstructed 
image with interstellar scattering (including both, diffractive and refractive scattering during the generation of the synthetic visibilities) using the space-EHT configuration convolved with 75\% (red shading) of the nominal beam size (light grey shading, 
panels c, f, i, l)} 
\label{finalimage} 
\vspace{-11pt} 
\end{figure*}

In order to quantify the ability to test 
different space-times we compute the structural dissimilarity measure (DSSIM) \citep{2004ITIP...13..600W} 
between the GRRT and reconstructed images. Before computing the DSSIM we 
perform an image alignment using the normalised cross correlation coefficient (nCCC) 
\citep[see, \eg][]{2018NatAs...2..585M}. Both image metrics are reported in the panels of Fig.~\ref{finalimage} and in 
Table~\ref{results}. In Table~\ref{results} we list the DSSIM and nCCC values computed for the 
different image combinations and for the different array configurations. If we could clearly 
distinguish {among} different theories based on the reconstructed images, the smallest DSSIM 
and largest nCCC values should be obtained for equal image pairs \eg boson star -- 
boson star (the diagonal in Table~\ref{results}). However, for the EHT 2017 configuration this 
is only true for the boson star. Thus based on reconstructed images obtained from synthetic data which take the source variability into account it is difficult for the EHT 2017 configuration to distinguish the spin of Kerr black holes and to differentiate a Kerr black from a dilaton black hole (see very similar values in 

\clearpage

Table \ref{results}). 
Including a space antenna 
improves the ability to distinguish {among} the different spacetimes, especially for the Kerr 
black hole with high spin. Given the obtained image metrics the space-EHT concept can distinguish the spin of the Kerr black holes, but it is still difficult to discriminate between a Kerr black hole with spin ${\rm a}_{*}$=0.6 and a dilaton black hole (see Table \ref{results}).

By comparing the image metrics obtained for EHT 2017 and the space-EHT concept two different behaviours can be found:  improved image metrics (decreased DSSIM and increased nCCC) and worsen ones (increased DSSIM and decreased nCCC). For equal image pairs  \eg dilaton-dilaton, the image metrics for the space-EHT concept improved as compared to the EHT 2017 (as discussed above). However for unequal image pairs there are two different behaviours. For example in the boson star - dilaton case the image metrics improved. In contrast to the dilaton - boson star pair where the image metrics become worse. This behaviour could be understood in the following way: The boson star image is very compact as compared to the dilaton model (see left column in Fig. \ref{finalimage}). Due to the limited resolution of the EHT 2017 array the intrinsically large structure of the dilaton model is smeared and scattered out to an even larger structure (compare panel i and j in Fig. \ref{finalimage}). However the improved imaging capabilities of the space-EHT concept leads to a more compact and less smeared out source structure which is reflected by a better matching between the boson star and the dilaton black hole. The contrary happens in the dilaton - boson star case: the EHT 2017 blurs the true boson star structure to a large size which matches better the dilaton structure as compared to sharper more compact boson star image provided by the space-EHT concept. As a result the image metrics for this image pair will become worse as compared to the EHT 2017. A similar behaviour can be seen for  Kerr ${\rm a}_{*}$=0.6 - boson star and to some extent for dilaton - Kerr ${\rm a}_{*}$=0.94 and Kerr ${\rm a}_{*}$=0.6 - Kerr ${\rm a}_{*}$=0.94. In Appendix \ref{dssimtest} we provide a more detailed study on the variation of the image metrics with respect to intrinsic source size and array resolution.

\begin{table}
\centering
\caption{Results of the cross-comparison of the synthetic images using the EHT 2017 and 
the advanced EHT configuration. The values correspond to the DSSIM and the numbers in 
brackets indicated the normalised cross correlation coefficient (nCCC). Small values for the 
DSSIM and numbers close to 1 for the nCCC indicate well-matched images (lowest DSSIM values are indicated in bold).}
\setlength{\tabcolsep}{0.3em} 
\begin{tabular}{@{}l l l l l @{}}
\hline\hline
& boson star & dilaton & Kerr (${\rm a}_{*}$=0.6) & Kerr (${\rm a}_{*}$=0.94)\\
\hline
\multicolumn{5}{c}{\emph{EHT 2017 configuration}}\\
\hline
boson star 	& \textbf{0.03 (0.98)} & 0.17 (0.79) & 0.18 (0.81) & 0.10 (0.88) \\
dilaton 		& 0.16 (0.83) & 0.13 (0.92) & {0.13 (0.92)} & \textbf{0.12 (0.93)} \\
Kerr (${\rm a}_{*}$=0.6)	& 0.16 (0.85) & 0.14 (0.91) & {0.13 (0.92)} & \textbf{0.12 (0.94)} \\
Kerr (${\rm a}_{*}$=0.94)	& \textbf{0.09 (0.89)} & 0.18 (0.83) & 0.18 (0.85) & \textbf{0.09 (0.93)} \\
\hline
\multicolumn{5}{c}{\emph{EHT including space antenna}}\\
\hline
boson star 	& \textbf{0.03 (0.98)} & 0.13 (0.88) & 0.14 (0.85) & 0.09 (0.89) \\
dilaton 		& 0.17 (0.78) & \textbf{0.10 (0.96)} & {0.11 (0.96)} & 0.12 (0.88) \\
Kerr (${\rm a}_{*}$=0.6)	& 0.17 (0.80) & 0.11 (0.96) & \textbf{0.10 (0.96)} & 0.12 (0.91) \\
Kerr (${\rm a}_{*}$=0.94)	& 0.09 (0.89) & 0.13 (0.89) & 0.13 (0.92) & \textbf{0.07 (0.96)} \\
\hline
\label{results}
\end{tabular}
\vspace{-16pt}  
\end{table} 

\subsection{Fourier Plane comparison}
\begin{figure}[h!]
\resizebox{8.8cm}{!}{\includegraphics{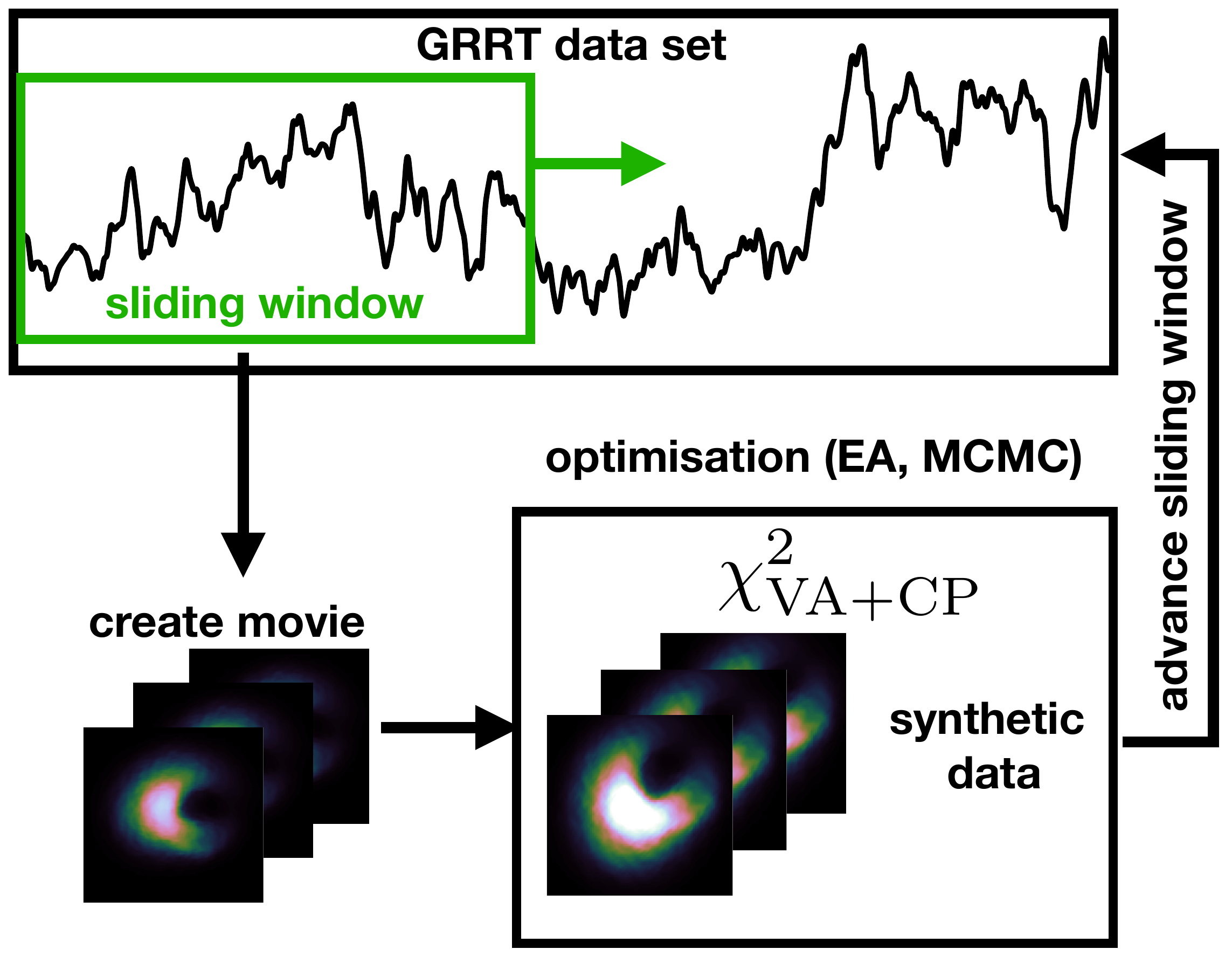}} 
\caption{Illustration of the movie scoring scheme. From the GRRT data we select via a sliding window a set of images from which a movie is created. From this movie complex visibilities are generated and the $\chi^2$ between the synthetic data is computed while allowing the movie to rotate and vary in source size and flux density. The $\chi^2$ are minimised either using an evolutionary algorithm (EA) or an MCMC scheme. After the optimisation the sliding window is advanced until the end of GRRT data set is reached.} 
\label{moviescoring} 
\end{figure}

{Given that an interferometer measures the Fourier transform of the brightness distribution 
of an astronomical source, a more direct comparison between images of different space-times can 
be obtained in Fourier space. The turbulent nature of the accretion process in the
GRMHD simulations manifests itself in large variations in the total flux density and in
its flux density distribution (see Fig.~\ref{fluxvar} for Kerr ${\rm a}_{*}$=0.6). 
In this work we include the variability of the source in the scoring procedure. 
Therefore we modify the scheme used in \citet{2019ApJ...875L...6E} for the analysis 
of the recent EHT M87 observations.
The main modification is that the for the comparison with the synthetic data we use a GRRT movie generated from a series of GRRT images spanning the observing time given by the synthetic data set. In this work typically 216 frames (for an EHT observation of 12\,h) are used and we slide along our GRRT data series for the different spacetimes where an increment of 5 frames is used. The increment of 5 frames is justified by the fact that the typical correlation times of the GRRT images {are} around 50\,M. This implies that around this time the GRRT images can be regarded as independent realisations of the accretion flow and thus the different movies created from the sliding window can be considered as uncorrelated. 
The first step in the scoring procedure is to create complex visibilities from the GRRT movie, taking into account the array configuration, the observing schedule and interstellar scattering, including both, diffractive and refractive scattering (see also Table~\ref{SEFD} and \ref{reconst}). From the complex visibilities we compute the visibility amplitude, VA, and from closed antenna triangles the closure phase, CP. The latter is of great importance in measuring the structural variation within the source. In the second step we minimise the $\chi^2$ for VA and CP between the snapshots and the individual frames (images) of the GRRT movie by allowing the total flux, the position angle and the black hole mass\footnote{Actually, we vary the plate scale, $\mu=m_{bh}/d_{bh}$, during the scoring. Assuming a fixed distance $d_{bh}$ to SgrA$^{*}$, we compute the black hole mass, $m_{bh}$, from $\mu$.} to vary. Notice that once the values for the flux scaling, position angle and black hole mass are set, they are kept constant for all frames of the GRRT movie in order to ensure the {consistency} of the movie. 
In addition to the image scaling, we perform antenna gain calibrations. We limit the variation in the individual antenna gains to be between 50\% and 150\% in order to avoid the compensation of structural differences {among} the models by large gain variations. The scoring is carried out with the well tested \textit{GENA} pipeline developed for the image matching of EHT and VLBA observations \citep{2019ApJ...875L...5E,2019ApJ...875L...6E,2019A&A...629A...4F} and we focus our analysis on the synthetic observations including a space-antenna.
The scheme for this kind of scoring which includes the source variability (hereafter  \textit{movie scoring}) is illustrated in Fig. \ref{moviescoring}.}

\begin{figure}[h!]
\resizebox{8.8cm}{!}{\includegraphics{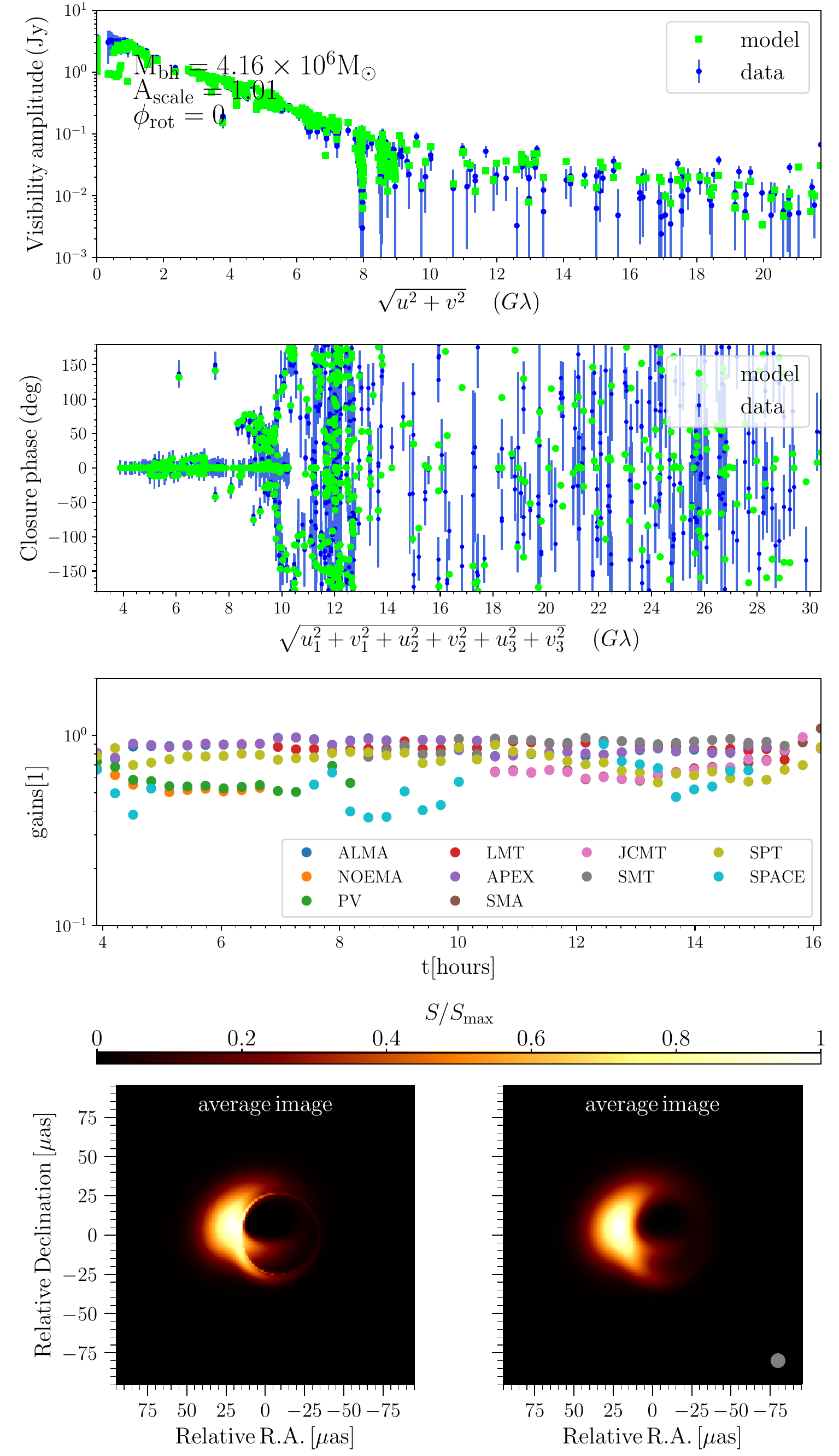}} 
\caption{Scoring result of the Kerr ${\rm a}_{*}$=0.94 synthetic data to the Kerr ${\rm a}_{*}$=0.94 model. The panels show from top to bottom the visibility amplitude, the closure phase, the calibrated gains and the average image of the Kerr ${\rm a}_{*}$=0.94 movie (GRRT left and convolved right).} 
\label{Kerr09scoring} 
\end{figure}

 {In Fig. \ref{Kerr09scoring} we show an example for movie scoring for Kerr ${\rm a}_{*}$=0.94 synthetic data  to the Kerr ${\rm a}_{*}$=0.94 model. The visibility amplitude is plotted in the top panel and the closure phase in the second panel. The blue points correspond to the synthetic data and green ones to the movie. In the third panel we present the calibrated gains for the antennas involved in the observations. 
The bottom panel shows the static average frame from the Kerr movie (GRRT image left and convolved on the right). The obtained values for the mass, the flux scaling and position angle are indicated in the top panel. The recovered values are in good agreement with the injected values. A more detailed self-test of the \textit{movie scoring} scheme is presented in the Appendix \ref{movietest}.\\  }
\begin{figure}[h!]
\resizebox{8.8cm}{!}{\includegraphics{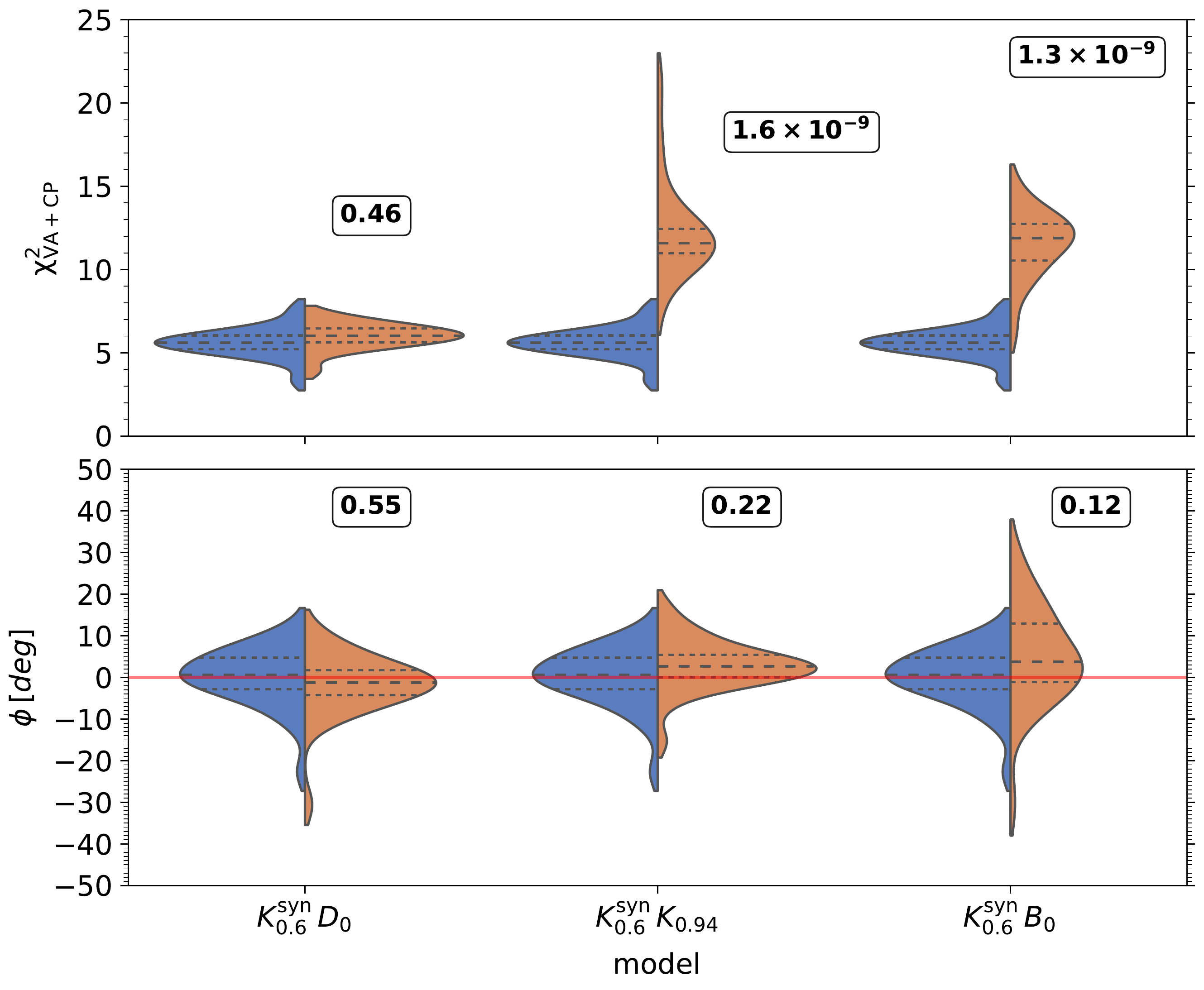}} 
\caption{Results for the Kerr ${\rm a}_{*}$=0.6 test. The panels show the distribution of total $\chi^2=(\chi^2_{VA}+\chi^2_{CP})/2$ (top) and the position angle, $\phi$ (bottom). In each violin the left hand side corresponds to ${K^{\rm syn}_{0.6}}-K_{0.6}$ and the numbers above the violins indicate the results of the two-sided K-S test (see text for further details). The red line in the bottom panel corresponds to the initial position angle, $\phi=0$ of the GRRT images used to create synthetic data for ${K^{\rm syn}_{0.6}}$.} 
\label{Kerr06vio} 
\end{figure}
\begin{figure}[h!]
\resizebox{8.8cm}{!}{\includegraphics{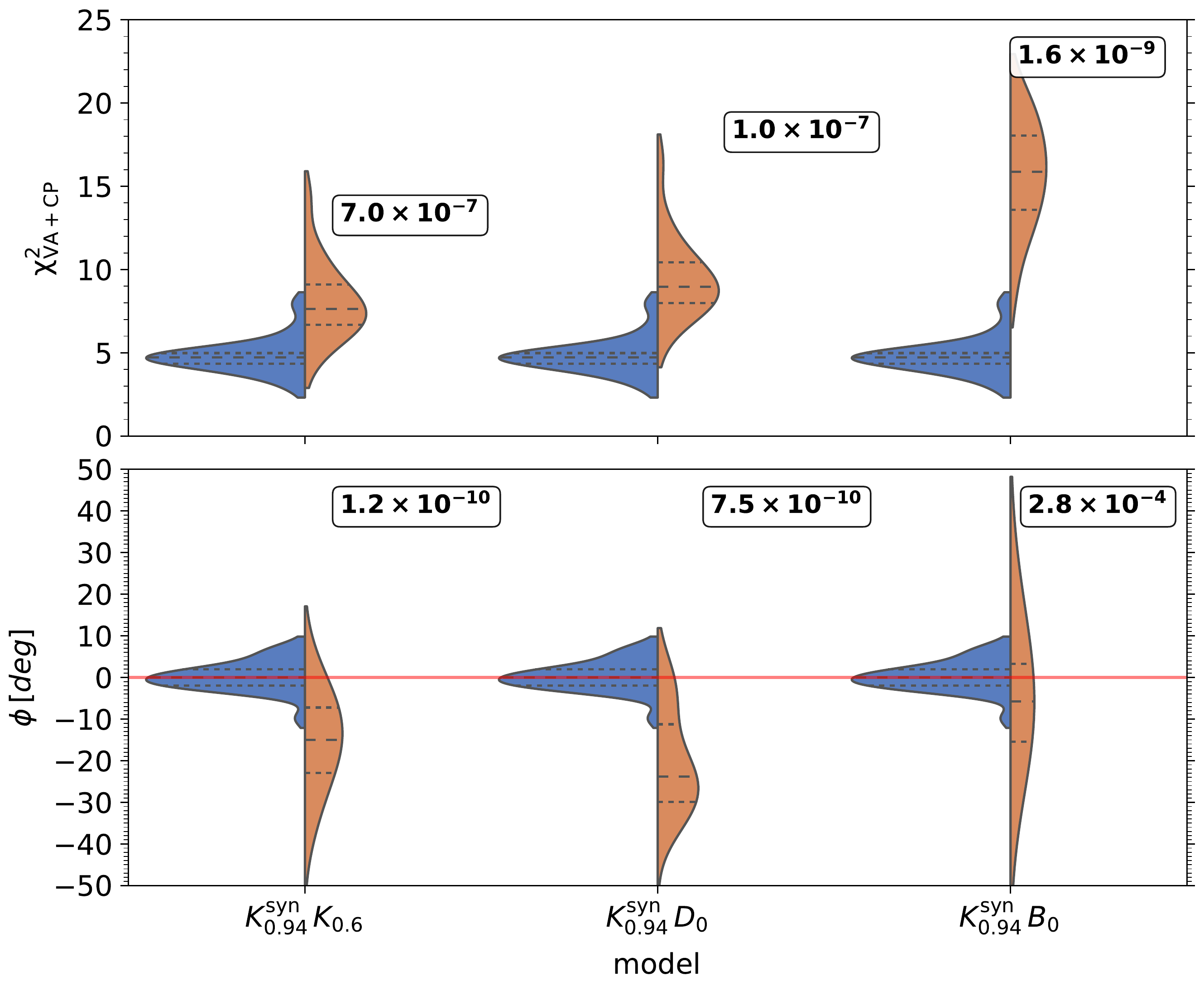}} 
\caption{As in Fig.~\ref{Kerr06vio}, now for the Kerr ${\rm a}_{*}$=0.94 test.} 
\label{Kerr09vio} 
\end{figure}
\begin{figure}[h!]
\resizebox{8.8cm}{!}{\includegraphics{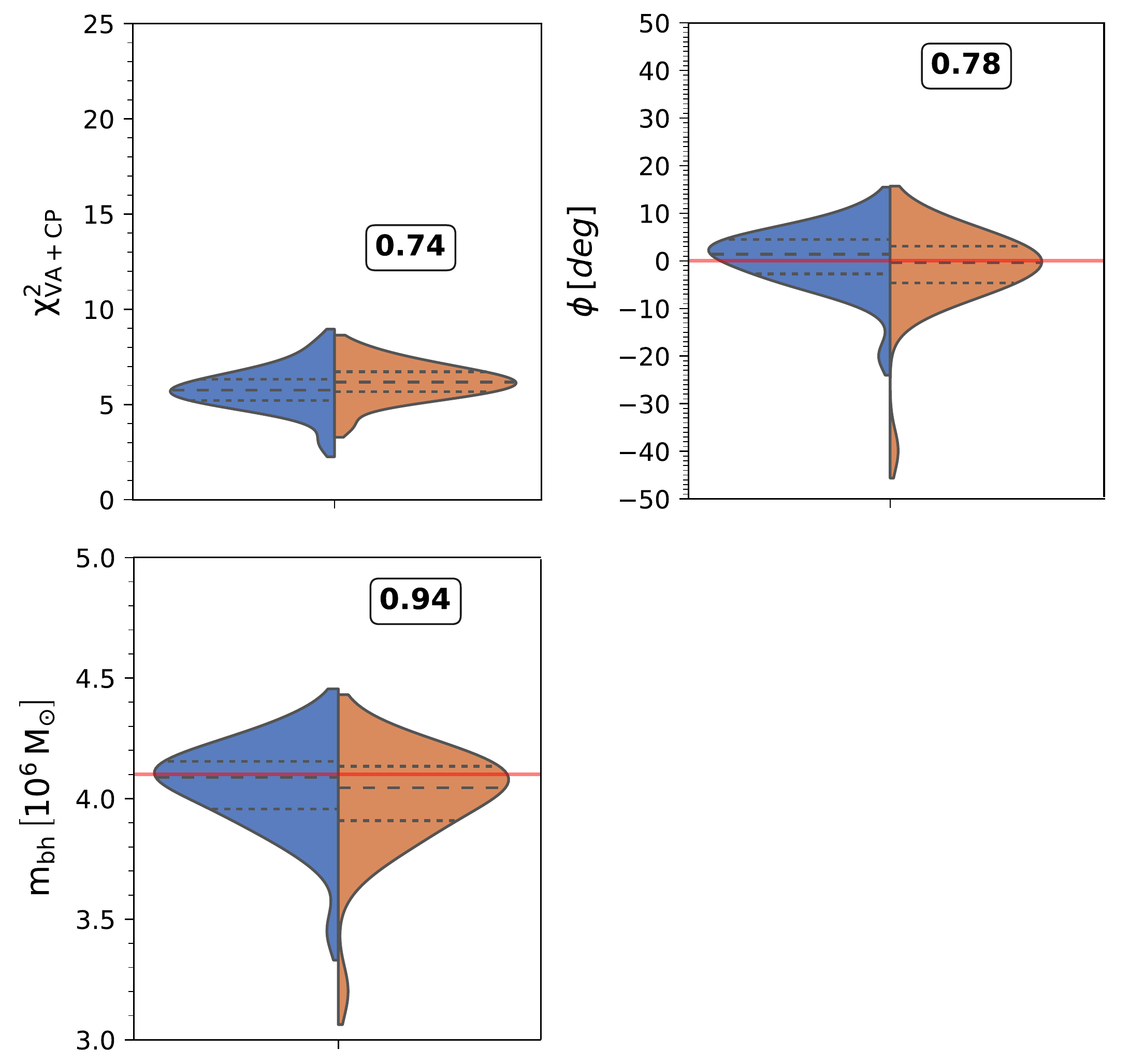}} 
\caption{Results for the Kerr ${\rm a}_{*}$=0.6 -dilaton black hole test with varying black hole mass. The panels show the distribution of total $\chi^2=(\chi^2_{VA}+\chi^2_{CP})/2$ (top left),  the position angle, $\phi$ (top right) and the black hole mass (bottom). In each violin the left hand side corresponds to ${K^{\rm syn}_{0.6}}-K_{0.6}$ and the right hand side to ${K^{\rm syn}_{0.6}}-D_{0}$. The numbers above the violins indicate the results of the two-sided K-S test. The red line in the middle and right panel corresponds to the initial position angle $\phi=0$ and initial black hole mass $M_{\rm bh}=4.14\times10^6M_{\astrosun}$ used for the GRRT images.} 
\label{Kerr06varymass} 
\end{figure}

{During our analysis we fit the synthetic data generated from the first 12\,h of our GRRT data set to the entire data by advancing the sliding window with an increment of 5 frames. This allows us to statistically quantify the difference between the synthetic data generated from the first 12\,h to the entire GRRT data set including the source variability. In the next step we keep the synthetic data and score it against the GRRT images of remaining three different spacetimes. Finally, we obtain the $\chi^2$-distribution for VA, CP, the position angle (measured from north through east) and the mass distribution for various combinations of synthetic data and the entire GRRT image sets for different spacetimes. Based on these distributions we perform a Kolmogorov--Smirnov (K-S) test to determine if the synthetic data under investigation is in agreement of being drawn from one of the four spacetimes. \\
Below we provide a short example which explains the generation of the violin plots and the obtained p-value from the K-S test.
Assume two normal distributions one with mean at 5.5 and standard deviation of 0.25 and another one with mean at 5.7 and standard deviation of 0.35. Now draw 20 random numbers for the the distribution and create a violin plot (the left half of the violin is the histogram for the first distributions and the right half corresponds to the second distribution). In this case the violin looks nearly {symmetrical} with only a slight shift and the computed K-S test provides a p-value of 0.49. This value implies that we cannot reject the hypothesis that the values are from the same ``overall'' distribution or in your case the underlying model (spacetime) is the same. (see left most violin in the top panel of Fig. \ref{Kerr06vio}).
On the contrary if the mean of the second distribution is located at 12 the two wings of the violin are disjoint and the K-S test provides a p-value of 5$\times10^{-10}$. This value indicates that the two distributions are not draw from the same ``overall'' distributions. In our case this would imply that {the} underlying model (space-time) is not the same (see right most violin in Fig. \ref{Kerr06vio}).}

In the following we label the different spacetimes by their first letter and a subscript indicates their spin, for example $K_{0.6}$ for a Kerr black hole with spin ${\rm a}_{*}$=0.6. We indicated the synthetic data by a superscript ``syn'' for example ${K^{\rm syn}_{0.6}}$ for the synthetic data of the Kerr black hole with spin ${\rm a}_{*}$=0.6 generated from the first 12\,h. Using this notation, scoring the synthetic data  Kerr ${\rm a}_{*}$=0.6 against the entire data set of Kerr ${\rm a}_{*}$=0.94 is given by ${K^{\rm syn}_{0.6}}-K_{0.94}$.

\subsubsection{Application to Kerr ${\rm a}_{*}$=0.6 (test 1)}
In a first application of the \textit{movie scoring}  technique we want to test whether we can distinguish the synthetic data for a Kerr black hole with spin ${\rm a}_{*}$=0.6 from a Kerr black hole with spin a$_{*}$=0.94, a dilaton black hole and from a boson star. Therefore, we score the synthetic data for Kerr black with spina$_{*}$=0.6 against the data set for the different space times and perform the KS test. Given the known black hole mass for the black hole in SgrA${^\star}$ and its small uncertainty from the GRAVITY experiment \citep{2019A&A...625L..10G} we keep for the black hole mass fixed for this first test and perform the K-S test only on the total $\chi^2=(\chi^2_{VA}+\chi^2_{CP})/2$ and the position angle. 

Figure \ref{Kerr06vio} shows the results for the Kerr ${\rm a}_{*}$=0.6 test. In the panels the left half of each violin (blue) corresponds to ${K^{\rm syn}_{0.6}}-K_{0.6}$ and the right half (from left to right) to ${K^{\rm syn}_{0.6}}-D_{0}$, ${K^{\rm syn}_{0.6}}-K_{0.94}$ and ${K^{\rm syn}_{0.6}}-B_{0}$, respectively.
The number above the violins indicates the results of the two-sided K-S test.
Within each half-violin the long-dashed lines indicate the mean and the short dashed line denotes the inter-quartile range.
The top panel displays the $\chi^2$ distribution, the bottom panel the distribution of the position angle, $\phi$.
The red line in the bottom panel indicates initial position angle of images which are used to create the synthetic data set.
For ${K_{0.6}}-D_{0}$ the $\chi^2$ show a very similar, only slightly shifted shape. In contrast to ${K_{0.6}}-K_{0.94}$ and ${K_{0.6}}-B_{0}$ where the distributions are only marginally overlapping. The distributions for the position angle $\phi$ are in all cases overlapping with the distribution of the truth model (${K^{\rm syn}_{0.6}}-K_{0.6}$, blue violins). \\

This behaviour can be explained by the very similar source size for the Kerr ${\rm a}_{*}=0.6$ and dilaton black hole in contrast to Kerr black hole with spin ${\rm a}_{*}=0.94$ and the boson star (see left column in Fig \ref{finalimage}). In order to improve the $\chi^2$ during the optimisation (or MCMC) step the image is rotated a bit more in the case of the Kerr ${\rm a}_{*}=0.94$ and the boson star data set. The results of the K-S test can be found in Table~\ref{ksresults}.\\
The results of the \textit{movie scoring} indicate that the space-EHT concept can clearly distinguish fast from slow spinning black holes as well as from boson stars. 

\subsubsection{Application to Kerr ${\rm a}_{*}$=0.94 (test 2)}
The second test we use the synthetic data from the  Kerr black hole with spin ${\rm a}_{*}$=0.94 and
we test if it can be distinguished from either a Kerr black hole with spin ${\rm a}_{*}$=0.6, a dilaton black hole or a boson star. 
The following movie scorings are computed: ${K^{\rm syn}_{0.94}}-K_{0.94}$ (synthetic data and GRRT data set are from the same spacetime),
${K^{\rm syn}_{0.94}}-D_{0}$, ${K^{\rm syn}_{0.94}}-K_{0.6}$ and ${K^{\rm syn}_{0.94}}-B_{0}$. The $\chi^2$ and position angle distribution of all test pairs of models are clearly shifted (see Fig.~\ref{Kerr09vio}). This behaviour is also reflected in the small numbers for the K-S test (see Table~\ref{ksresults}).The shift in the $\chi^2$ distributions can be explained by the difference in source size for the Kerr ${\rm a}_{*}$=0.6 and dilaton black hole. In order to match the source size of the Kerr black hole with spin ${\rm a}_{*}$=0.94 the dilaton and Kerr black hole with spin ${\rm a}_{*}$=0.6 would require a smaller black hole mass $m_{\rm bh}<4.14\times10^6\, M_{\astrosun}$. Since we do not allow the black hole mass to adjust the only way to improve the $\chi^2$ is to rotate the GRRT movies. An improved $\chi^2$ is obtained by rotating the GRRT movies were the mean of the $\phi$ distributions is located at -15$^\circ$ ( Kerr black hole with spin ${\rm a}_{*}$=0.6) and -25$^\circ$ (dilaton black hole). Similar, the boson star would require a larger black hole mass to fit the Kerr ${\rm a}_{*}$=0.94 synthetic data. \\
Given the distributions and the result of the K-S test the space-EHT concept can distinguish {among} all three models if the truth model is a Kerr black hole with spin ${\rm a}_{*}$=0.94.
{\subsubsection{Application to Kerr ${\rm a}_{*}$=0.6 with varying black hole mass (test 3)}
In order to test the variations in the black hole mass required to make the Kerr ${\rm a}_{*}$=0.6 indistinguishable for the space-EHT concept from the dilaton black hole we performed a third test where we restrict ourselves to Kerr ${\rm a}_{*}$=0.6 and dilaton black holes and allow the black hole mass to adjust during the optimisation. The result for this test is presented in Fig. \ref{Kerr06varymass}. As expected the adjusted black hole mass for the dilaton black hole improved the p-values for the K-S test on all quantities ($\chi^2_{\rm tot}$, $\phi$ and $m_{\rm bh}$). However this would {require} a black hole mass of $m_{\rm bh}=4.05\times10^6\, M_{\astrosun}$ which is not in agreement with the current measurements of GRAVITY \citep{2019A&A...625L..10G}.}

\begin{table}[h!]
\renewcommand{\arraystretch}{1.3}
\centering
\caption{Results of the K-S tests for Kerr ${\rm a}_{*}$=0.6 and Kerr ${\rm a}_{*}$=0.9 as truth images. The K-S tests are performed
on the distribution of the total $\chi^2_{\rm VA+CP}$ (visibility amplitude and closure phases) and the distribution of the black hole mass, $m_{\rm bh}$.}
\setlength{\tabcolsep}{0.5em} 
\begin{tabular}{@{}l l l l @{}}
\hline\hline
\multicolumn{4}{c}{truth model ${K^{\rm syn}_{0.6}}$ (test 1)}\\
\hline
&  ${K^{\rm syn}_{0.6}}-D_{0}$ & ${K^{\rm syn}_{0.6}}-K_{0.94}$ & ${K^{\rm syn}_{0.6}}-B_{0}$\\
\hline
p-value $\left(\chi^2 \right)$	&$0.46$ & $1.6\times10^{-9}$ & $1.3\times10^{-9}$  \\
p-value $\left(\phi \right)$		&$0.55$ & $0.22$ & $0.12$ \\
\hline\hline
\multicolumn{4}{c}{truth model ${K^{\rm syn}_{0.94}}$ (test 2)}\\
\hline
& ${K^{\rm syn}_{0.94}}-D_{0}$ & ${K^{\rm syn}_{0.94}}-K_{0.6}$ & ${K^{\rm syn}_{0.94}}-B_{0}$ \\
\hline
p-value $\left(\chi^2 \right)$ &  $7.0\times10^{-7}$ & $1.0\times10^{-7}$ & $1.6\times10^{-9}$\\
p-value $\left(\phi \right)$	& $1.2\times10^{-10}$ & $7.5\times10^{-10}$ & $2.8\times10^{-4}$\\
\hline \hline
\multicolumn{4}{c}{truth model ${K^{\rm syn}_{0.6}}$ (test 3)}\\
\hline 
& \multicolumn{3}{c}{${K^{\rm syn}_{0.6}}-D_{0}$ }\\
\hline
p-value $\left(\chi^2 \right)$	&  \multicolumn{3}{c}{0.74}\\
p-value $\left(\phi \right)$		& \multicolumn{3}{c}{0.78}\\
p-value $\left(m_{\rm bh}\right)$ & \multicolumn{3}{c}{0.94}\\
\hline
\label{ksresults}
\end{tabular} 
\vspace{-16pt}  
\end{table} 
\section{Discussion and Summary}
\label{dis}
{In this exploratory paper we address the question if an orbiting space antenna will improve the capabilities of the EHT to distinguish among different space times around SgrA$^\star$. Our proposed optimisation procedure suggested an elliptical orbit with an eccentricity of e=0.5 and a semi-major axis of a=14900\,km. By  including the suggested space antenna the baselines of the EHT array could be extended beyond 
10,000\,km and thus increase the angular resolution and the imaging capabilities of the array considerably (see right column in Fig. \ref{finalimage}).}
As can be seen in Table~\ref{results} the space-EHT is able to distinguish the spin of Kerr black holes (best image metrics can be found for equal image pairs). For example the DSSIM for the 
Kerr ${\rm a}_{*}$=0.6 -- Kerr ${\rm a}_{*}$=0.6 dropped from 0.13 to 0.10 while the Kerr ${\rm a}_{*}$=0.6 -- Kerr ${\rm a}_{*}$=0.94 is 0.12. The improvement on the imaging capabilities of the space-EHT is also noticeable in the change of the computed nCCC: for Kerr ${\rm a}_{*}$=0.6 from 0.91 to 0.96 while for the Kerr ${\rm a}_{*}$=0.94 case the nCCC decreased from 0.94 to 0.91. A similar behaviour is found for the other equal image pairs \eg Kerr ${\rm a}_{*}$=0.94 -- Kerr ${\rm a}_{*}$=0.94 as compared to the non-equal image pairs \eg Kerr ${\rm a}_{*}$=0.94 -- boson star (see last row in Table~\ref{results}).
For the ${\rm a}_{*}$=0.6 Kerr black hole and the dilaton black hole the image metrics improved on very similar: the DSSIM value: from 0.13 to 0.10 and the nCCC from 0.92 to 0.96. Given these very similar number is currently not possible to distinguish both spacetimes based solely on their reconstructed images.

To circumvent the limitations of reconstructed average images in distinguishing different theories of gravity, we additionally
performed a detailed comparison in Fourier space including the source variability. 

{We created synthetic data from the first 12\,h of the GRRT images 
of the two different Kerr black holes and scored them against 12\,h movies creates from the GRRT images of all four spacetimes.
Fitting the synthetic data for the Kerr black hole with spin  ${\rm a}_{*}$=0.6 against its entire GRRT data set leads to a $\chi^2$ distribution
with a mean value of 5.6 and standard deviation of 0.5 (see blue violins in Fig~\ref{Kerr06vio}).
The position angle distribution peaks at a value of $\phi=0.7^\circ\pm4^\circ$.
This value is close the initial value we used for the creation of the synthetic data and can be regarded
as confirmation of the fitting routine. }
 
{We performed two-sided K-S tests to investigate the hypothesis that the synthetic data is drawn from a
spacetime other than Kerr.
The result for the Kerr ${\rm a}_{*}$=0.6 tests reveal very small numbers for the synthetic data generated from
Kerr ${\rm a}_{*}$=0.9 and a boson star, both for the $\chi^2$ and for the position angle
(see Fig~\ref{Kerr06vio} and Table~\ref{ksresults}).
Thus we could reject the null-hypothesis and conclude that Kerr ${\rm a}_{*}$=0.9 and a boson star can be
distinguished from an ${\rm a}_{*}$=0.6 Kerr black hole.
The K-S test between the ${\rm a}_{*}$=0.6 Kerr and the dilaton black holes provides values significantly larger
than for the Kerr ${\rm a}_{*}$=0.94 and a Boson star.
The mean values of the dilaton distributions are shifted by roughly one standard deviation compared to
the truth distribution (Kerr ${\rm a}_{*}$=0.6).
Given the obtained values, it is likely (54\% level) that we can distinguish an ${\rm a}_{*}$=0.6 Kerr
black hole from a dilaton black hole. In an additional test we allowed the black hole mass to adjust and in order to probe the mass variation which would lead to undistinguishable data sets \eg we can not differentiate between a Kerr black hole with spin ${\rm a}_{*}$=0.6 and a dilaton black hole. The shift between the $\chi^2$ and $\phi$ distribution decreased which leads to larger p-values for the K-S test (see Table \ref{ksresults} and Fig. \ref{Kerr06varymass}). In addition to the previous test we also obtain the distribution for the black hole mass. The obtained mean black hole mass for Kerr black hole with spin ${\rm a}_{*}$=0.6 is $m_{\rm bh}=4.1^{+0.05}_{-0.15}\times10^6\, M_{\astrosun}$ which is in good agreement with used black hole mass during the radiative transport calculations and a confirmation of the \textit{movie scoring} method. In order to make both theories of gravity undistinguishable to 74\% the mean black hole mass should decrease to  $m_{\rm bh}=4.02\times10^6\, M_{\astrosun}$. However, given the mass boundaries from the GRAVITY experiment this black hole mass is outside the allowed range. 
Similar to the ${\rm a}_{*}$=0.6 Kerr case, we perform K-S tests on the fast spinning Kerr black hole with
${\rm a}_{*}$=0.94.
Again, we obtain the distribution for the $\chi^2$ and the position angle $\phi$ and computed the K-S test.
All distributions for this test are clearly offset from the truth distribution (see top panel in Fig.~\ref{Kerr09vio}),
which is also reflected in the very small numbers obtained for the p-value (see Table~\ref{ksresults}).
However, given the small p-values we conclude that the fast-spinning Kerr black hole can be clearly distinguished
from the slower-spinning Kerr black hole, the dilaton and the boson star.}

{ In summary, we have presented a future EHT concept which includes a space antenna and tested the capabilities
of this new array to investigate the different possible spacetimes around Sgr A$^{*}$ via radio images and complex
visibilities.
The orbit of the satellite is computed from a non-linear optimisation using a genetic algorithm and constraints on
the {u-v-plane} filling, observation time and improved image metrics computed between the GRRT image and the reconstructed image.
We generated synthetic data for three different theories of gravity, namely: a Kerr black hole (general relativity),
a dilaton black hole and a boson star.
The synthetic data generated was created from 12\,h of GRRT movie in order to properly include the source variabily and a dynamical image reconstruction was applied following \citet{2017ApJ...850..172J} using \textit{ehtim}. From the dynamical reconstructed images an average image was computed and compared to the average frame of the GRRT movies using DSSIM and nCCC metrics.
The image plane comparison was accompanied by a more detailed and robust analysis in the Fourier plane
using the newly developed \textit{movie scoring} method in \textit{GENA}.
A K-S test was performed on the $\chi^2$ position angle and mass distributions, in order to investigate the possibility of distinguishing {among} different spacetimes and therefore different theories of gravity.
The space-EHT concept presented in this study has been shown to both improve the imaging capabilities while including the source variability  of the array and improve our ability to distinguish {among} (and potentially exclude) certain solutions and theories of gravity.\\}
{Our ability to image time-variable sources and probe the underlying spacetime will be further improved by the addition of additional ground antennas for example the African Millimetre Telescope (ATM) \citep{2016heas.confE..29B} or several telescopes placed at dedicated locations across the globe as planed by the next generation EHT (ngEHT) \citep{2019arXiv190901411B}. Extending the GRRT images series will allow {us} to address the source variability on larger {non-overlapping} time windows improving the statistics. Increasing the observing frequency ($\nu>500$\,GHz) will  reduce the interstellar scattering {and} also allow us to study more deeply general relativistic effects on horizon scales since the obtained image is less effected by the properties {of} the accretion flow and the radiation microphysics. However, due to the opacity of the Earths atmosphere such a concept would require an entirely space-based VLBI concept \citep[see \eg][]{2019A&A...625A.124R, 2019arXiv191013236V}. 
\newline In addition to horizon scale images of the EHT and future space-EHT concepts further constraints on the properties of the spacetime around Sgr A$^\star$ can be obtained from a pulsar in a tight orbit (orbital period $\sim1$\,yr ) around the galactic centre \citep{Liu_2012}. Combined, both measurements \eg horizon scale images and pulsar timing would allow tightly constraints of the properties of spacetime around SgrA$^\star$ \citep{2016ApJ...818..121P}.}
\begin{acknowledgements}
CMF thanks E. Ros, R. Porcas and D. Palumbo for fruitful discussion and comments on the manuscript.
Support comes from the ERC Synergy Grant ``BlackHoleCam - Imaging the
Event Horizon of Black Holes'' (Grant 610058).  CMF is supported by the black hole Initiative at Harvard University, which is supported by a grant from the John Templeton Foundation.
ZY acknowledges support from a Leverhulme Trust Early Career Fellowship.
\end{acknowledgements}
\bibliographystyle{aa} 
\bibliography{biblo_RHD}
\begin{appendix}

\section{Influence of the orbital parameters on the imaging capabilities}
\label{orbittest}
\begin{figure*}[h!]
\centering
\resizebox{0.95\hsize}{!}{\includegraphics{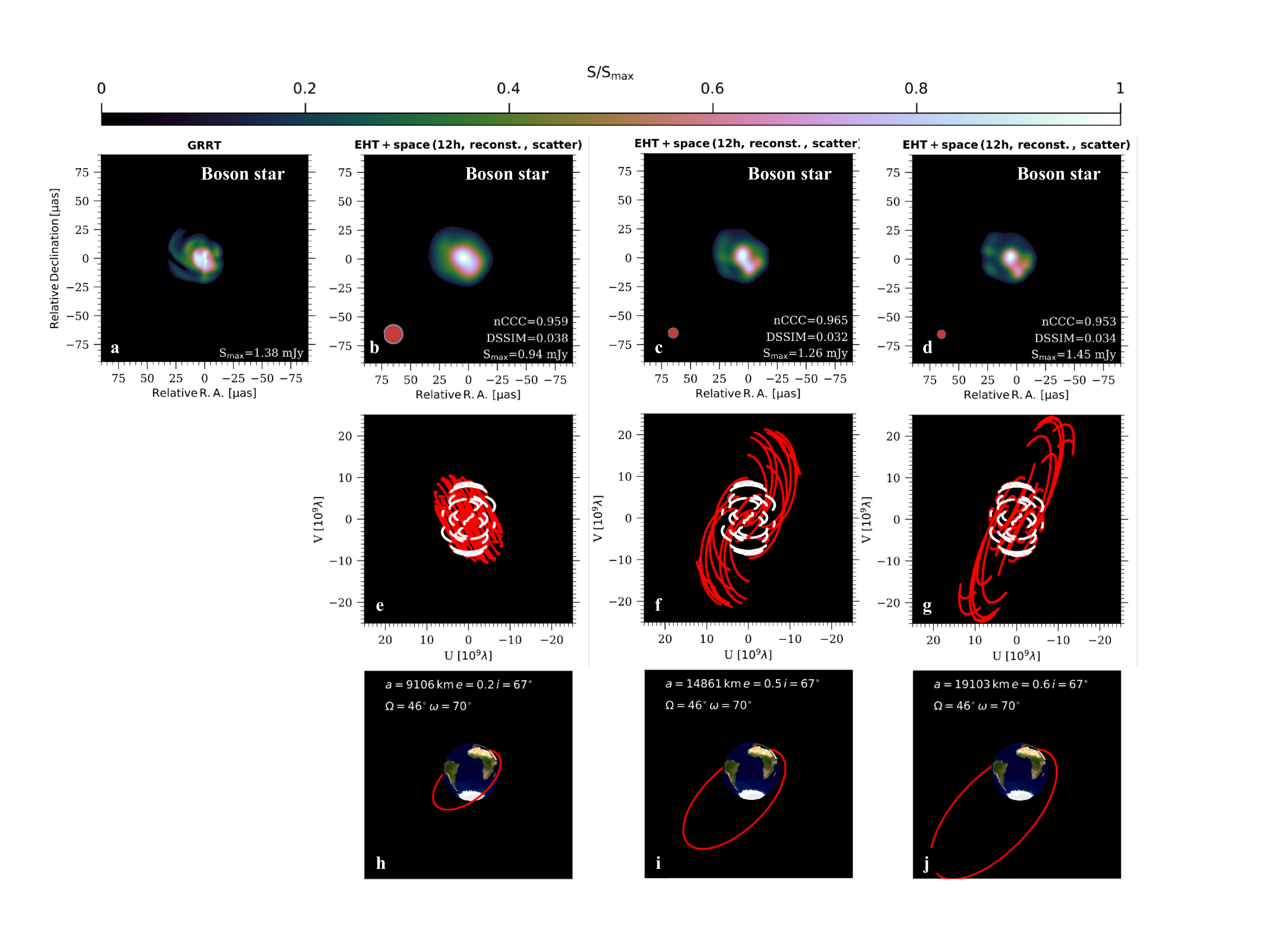}} 
\caption{Reconstruction of a boson star image for SgrA$^\star$ (ground truth image in panel a) for three different satellite orbits; small orbit (second column), intermediate orbit (third column) and large orbit (fourth column). The columns show from top to bottom the reconstructed image convolved with 75\% of the nominal array resolution, the uv-coverage (white: ground array baselines, red: space baselines) and the satellite orbit as seen from SgrA$^\star$.} 
\label{orbitexample} 
\end{figure*}

{In Fig. \ref{orbitexample} we illustrate the impact of the orbital parameters on the imaging capabilities of the space-EHT concept and the tradeoffs between u-v coverage and array resolution. We keep the orientation parameters of the orbit fixed to the results of our orbit optimisation (see Table \ref{resobs}) and change the semi-major axis, $a$, and eccentricity, $e$. }

{ An increase in the semi-major axis, $a$ (see panel j in Fig. \ref{orbitexample}), leads to higher resolution (see beam in panel d in Fig. \ref{orbitexample}). At the same time the larger semi-major axis, $a$,  which corresponds to a larger orbital period $T_p=2\pi\sqrt{a^3/\left(GM_{\rm Earth}\right)}$ reduces the u-v coverage (see panel g in in Fig. \ref{orbitexample}) . This leads to a worse image metric as for example an medium size orbit (compare image metrics in panel c and d in Fig. \ref{orbitexample}). On the other hand, a small semi-major axis (see panel h in Fig. \ref{orbitexample}) leads to a dense u-v coverage (panel e in Fig. \ref{orbitexample}) but does not provide high angular resolution (see panel b in Fig. \ref{orbitexample}). A trade-off between resolution and u-v coverage is an intermediate size semi-major axis (see third column in in Fig. \ref{orbitexample}). For each of the three satellite orbits the image metrics are calculated which prioritise the orbit with the best image metrics here the orbit with a semi-major axis of $a=14861$\,km and an eccentricity, $e=0.5$.}

\section{DSSIM and nCCC}
\label{dssimtest}
{In order to explore the variation of the DSSIM and the nCCC we performe a small parameter space study using averaged GRRT images.
The GRRT images are taken from Kerr BHs with spin ${\rm a}_{*}$=0.6 and ${\rm a}_{*}$=0.94 as well as from the dilaton black hole.
To mimic different observing arrays with improved imaging capabilities we convolve the average images with different beam sizes\footnote{we applied a circular gaussian} ranging from 30\,$\mu$as to 1\,$\mu$as. The convolved images are compared to the ground truth GRRT images and the DSSIM and the nCCC values are computed.}
\begin{figure}[h!]
\centering
\vspace{-10pt}
\resizebox{1\hsize}{!}{\includegraphics[]{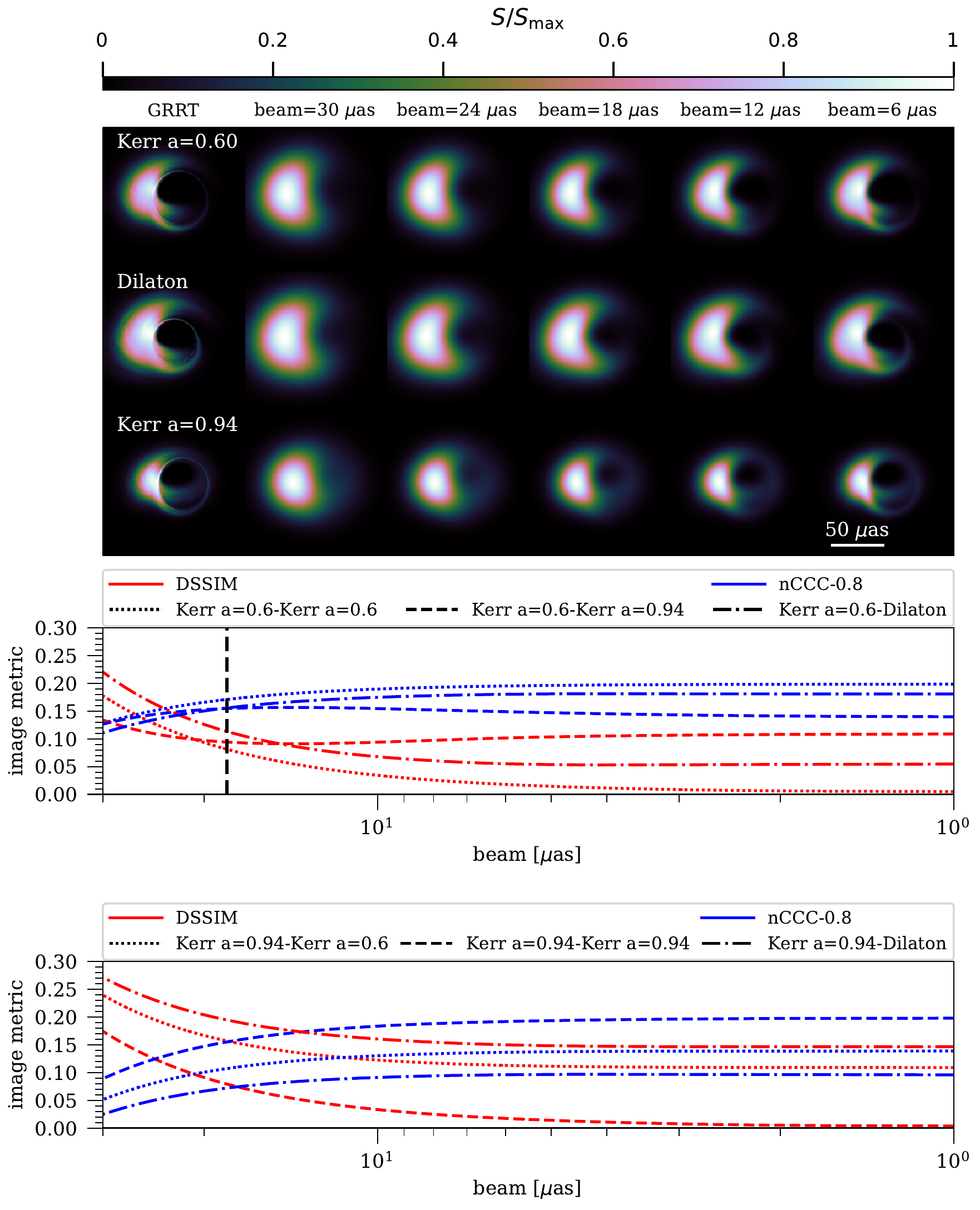}} 
\caption{DSSIM and nCCC variation study. Top panel: ground truth images for Kerr ${\rm a}_{*}$=0.6, Kerr ${\rm a}_{*}$=0.94 and dilaton black holes (left most column) and convolved with decreasing beams (30\,$\mu$as to 1\,$\mu$as). Middle panel: DSSIM and nCCC calculations using the average image of Kerr BH with ${\rm a}_{*}$=0.6 as ground truth image. Bottom panel: DSSIM and nCCC calculations using the average image of Kerr BH with ${\rm a}_{*}$=0.94 as ground truth image.} 
\label{dssim} 
\vspace{-11pt} 
\end{figure}
{In Fig. \ref{dssim} we show the results for this study. In the top panel we display the average GRRT images for the Kerr black holes (first and second row) and for the dilaton black hole (third row). Note that the images are individually normalised by their flux density maximum. The first column presents the ground truth GRRT image followed by a series of convolved images using beams with decreasing size (beam size is indicated above each image). With decreasing beam size (equal to the improved imaging capabilities of the array) the images become sharper and finer image features become visible. In the second panel of Fig. \ref{dssim} we compute the image metrics assuming that the ground truth image corresponds to a Kerr BH with spin ${\rm a}_{*}$=0.6. In the panel the red colour corresponds to the DSSIM and blue to the nCCC while the line style indicates the image pairs, \eg red dashed line stands for the DSSIM computed between Kerr BH with ${\rm a}_{*}$=0.6 and Kerr BH ${\rm a}_{*}$=0.94. As expected the DSSIM values decrease with smaller beam sizes while the nCCC increases for all image pairs. However at a beam size of $\sim 16\,\mu$as the DSSIM values for the Kerr BH with ${\rm a}_{*}$=0.94 increases and at the same time the nCCC decreases. This behaviour can be understood by the different source sizes of the ground truth images (see left most column in the top panel). The compact source structure of the Kerr BH with ${\rm a}_{*}$=0.94 is smeared out to larger extents during the convolution with beam with size $>16\,\mu$as which matches better the ground truth structure of the Kerr BH with ${\rm a}_{*}$=0.6. Thus, the DSSIM is decreasing while the nCCC is increasing. However at a beam size of $\sim 16\,\mu$as the behaviour is inverted. For the models with similar ground truth sizes the DSSIM is monotonically decreasing while the nCCC is continuously increases. \newline
If the size of the ground truth image is smaller than the intrinsic size of the models to which it is compared, \eg Kerr BH with ${\rm a}_{*}$=0.94 and Kerr BH ${\rm a}_{*}$=0.6 no change in the DSSIM and nCCC behaviour is found: the DSSIM is always decreasing and the nCCC is continuously increasing with decreasing beam size (see bottom panel in Fig. \ref{dssim}).}

\section{Movie scoring self-test}
\label{movietest}
{To valid the developed movie scoring method we perform a self-test. For the self-test we generate synthetic visibilities from the Kerr black hole with spin ${\rm a}_{*}$=0.94. The start frame for the synthetic data generation is shifted by 1.7 hours (30 frames). During the generation of the synthetic data we take interstellar scattering (including both, diffractive and refractive scattering), thermal noise and gain variations into account (see also Table \ref{reconst}). For a successful self-test the movie scoring method should recover the correct starting frame, the black hole mass and the orientation of the images used during the synthetic data generation. We during the radiative transfer we use  a black hole mass of $4.14\times 10^6\,M_{\astrosun}$ and an orientation angle of $\phi=0^\circ$. 
During the self-test we allow the black hole mass, $m_{\rm bh}$, the rotation angle, $\phi$ and the flux normalisation, S$_{\rm}$ to adjust during the scoring. For the optimisation we apply a MCMC method \citep{2013PASP..125..306F,corner} using 100 MCMC walkers for 500 iterations and perform gain-calibration.}

\begin{figure}[h!]
\centering
\vspace{-10pt}
\resizebox{1\hsize}{!}{\includegraphics[]{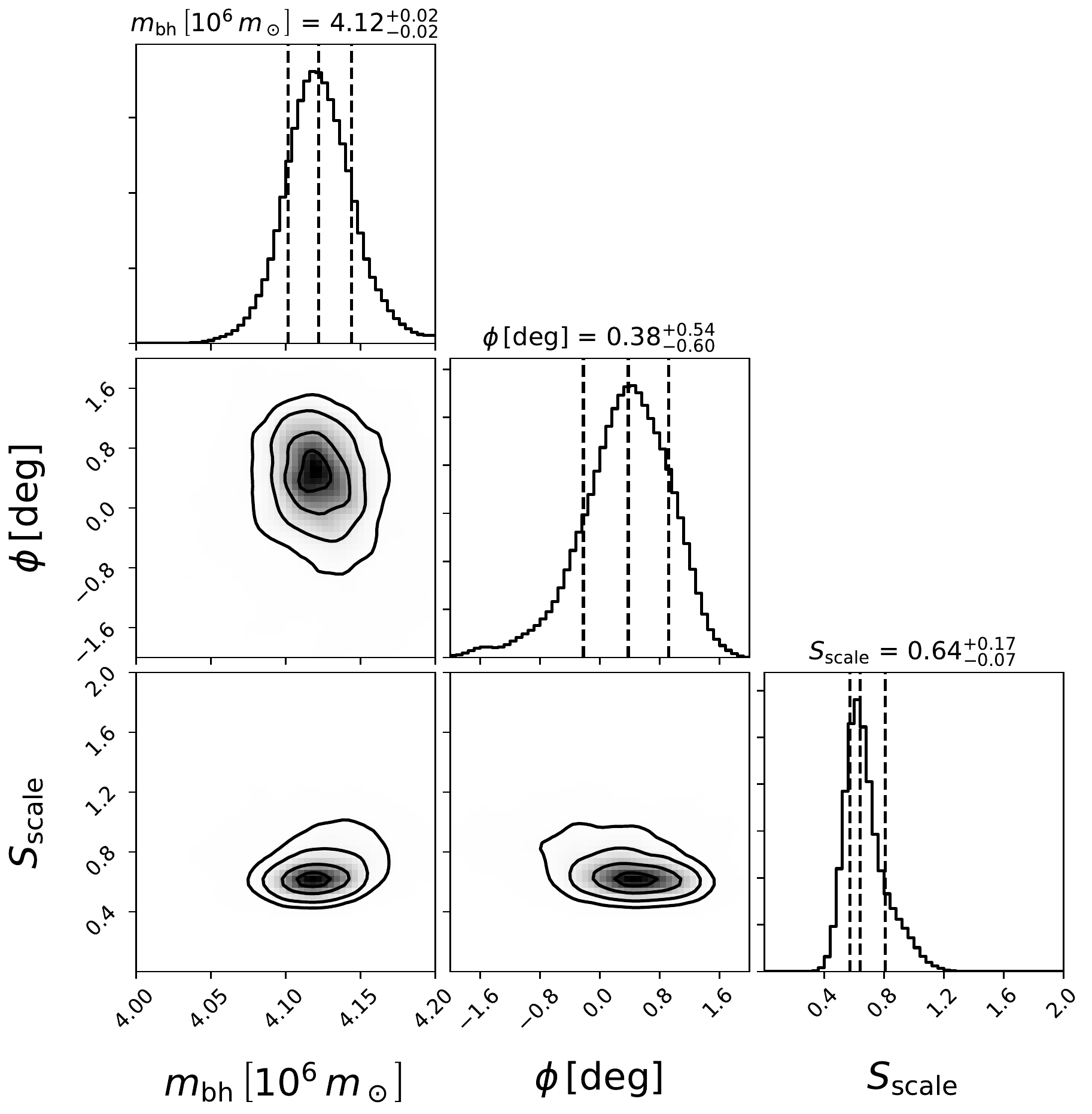}} 
\caption{Posterior distributions for the black hole mass, $m_{\rm bh}$, the position angle, $\phi$, and the flux scaling, $S_{\rm scale}$. The sliding window is located at $\Delta t=1.7$\,h (30 frames) recovering the time range used for the synthetic data generation (see Text).} 
\label{k094frame} 
\vspace{-11pt} 
\end{figure}

\begin{figure}[h!]
\centering
\vspace{-10pt}
\resizebox{1\hsize}{!}{\includegraphics[]{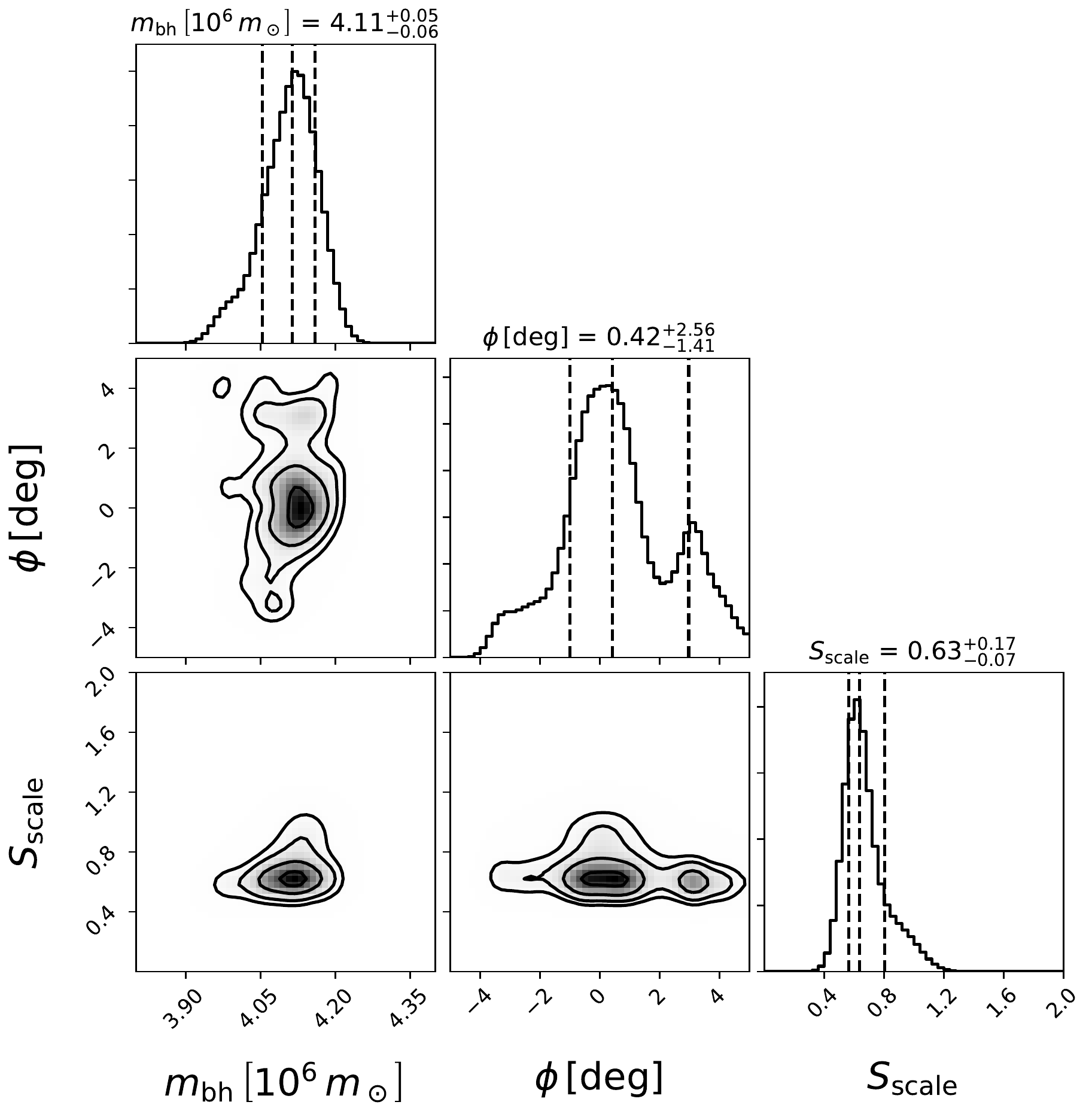}} 
\caption{Posterior distributions for the black hole mass, $m_{\rm bh}$, the position angle, $\phi$, and the flux scaling, $S_{\rm scale}$ for  the entire Kerr black hole ${\rm a}_{*}$=0.94 data set.} 
\label{k094all} 
\end{figure}

\begin{figure}[h!]
\centering
\vspace{-10pt}
\resizebox{1\hsize}{!}{\includegraphics[]{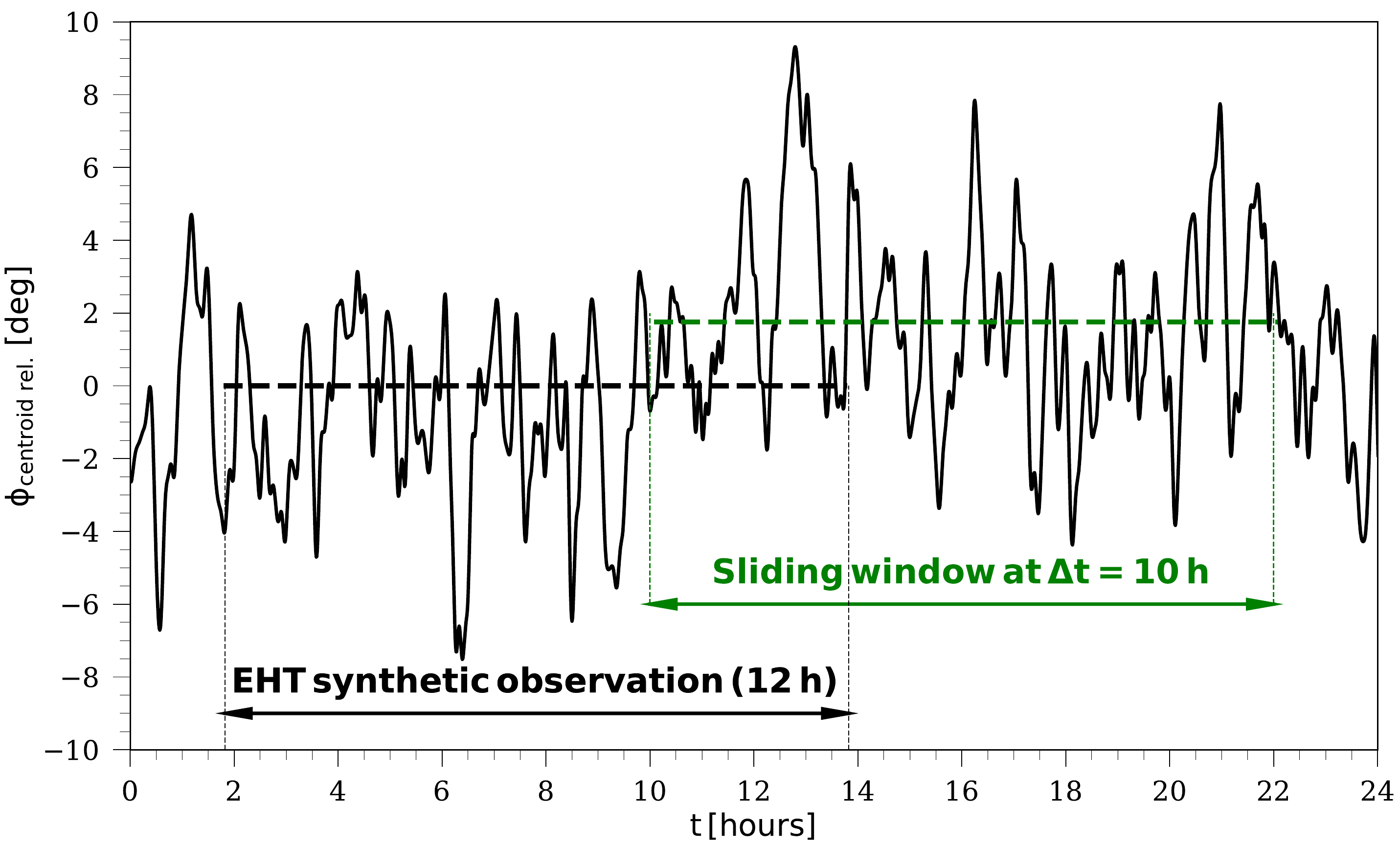}} 
\caption{Variation of the image centroid position angle relative to the one from the sliding window used for the generation of the synthetic data (indicated by the black arrow). The mean centroid position angle for a second sliding window with 10 hour offset is indicated in green.} 
\label{centroidpos} 
\vspace{-11pt} 
\end{figure}

{The movie scoring method successfully recovered the start frame and in Fig. \ref{k094frame} we present the posterior distribution for the black hole mass, the orientation angle and the flux scaling for the GRRT movie starting at frame 30. The black hole mass and the orientation angle are recovered within $1\sigma$ proofing the applicability of the developed method to recover black hole parameters from a source with varying flux density and structure. The posterior distribution of the flux scaling, $S_{\rm scale}$, reflects the influence of interstellar scattering and antenna gain variations taken into account during the synthetic data generation. Interstellar scattering smears out the flux density, thus reducing the measured visibility amplitude as compared to an un-scattered case. In order to match the visibility amplitude during the scoring the flux scaling factor is smaller than one.  The additional variations due to the antenna gain variations are compensated by the applied self-calibration. In addition we show in Fig. \ref{k094all} the posterior distribution for the scaling parameters ($m_{\rm bh}$, $\phi$ and $S_{\rm scale}$) obtained from the entire data set for the Kerr black with ${\rm a}_{*}$=0.94. The posterior distribution for the black hole mass, $m_{\rm bh}$, and the flux scaling, $S_{\rm scale}$ show a similar distribution as for the best starting frame model (see Fig. \ref{k094frame}) except for the three times larger uncertainties for the black hole mass. However the distribution of the position angle, $\phi$, shows a second local maximum around $\phi=3^\circ$. This can be explained by the variability of the source or more precise the variation of the position angle of the flux centroid. In Fig. \ref{centroidpos} we show the variation of the image centroid position angle relative to the data set used for the synthetic data. For sliding windows different that the one used for the data generation (indicated by the black arrow in Fig. \ref{centroidpos}) the centroid position angle is lager angle (see green dashed line in Fig.\ref{centroidpos}). During the optimisation within the movie scoring method this angle difference is compensated by a rotating the images  which explains the second maximum in the position angle posterior distribution in Fig. \ref{k094all}.}

\end{appendix}
\end{document}